\documentclass[final,5p,times,twocolumn,nopreprintline]{elsarticle}
\biboptions{sort&compress}
\usepackage[T1]{fontenc}
\usepackage[export]{adjustbox}
\usepackage[english]{babel}
\usepackage[utf8]{inputenc}
\usepackage{color,graphicx}
\usepackage{bm,bbm}
\usepackage{enumerate}
\usepackage{hyperref}
\usepackage{mathtools}
\usepackage{mathrsfs}
\usepackage{mciteplus}
\usepackage{slashed}
\usepackage{soul}
\usepackage[normalem]{ulem}
\usepackage{wrapfig}
\usepackage{xspace}
\usepackage[shortlabels]{enumitem}
\usepackage{orcidlink}
\usepackage[capitalise]{cleveref}
\usepackage{txfonts}  
\usepackage{relsize}
\usepackage{multirow}
\usepackage{gensymb}

\newcommand{\bcol}{\left[ \begin{array}{c}}
\newcommand{\ecol}{\end{array} \right]}
\newcommand{\beq}{\begin{eqnarray}}
\newcommand{\eeq}{\end{eqnarray}}

\newcommand{\ie}{{i.e.}\xspace}
\newcommand{\HD}{{$H$-dibaryon}\xspace}

\newcommand{\kev}{\ensuremath{{\mathrm{\,ke\kern -0.1em V}}}\xspace}
\newcommand{\mev}{\ensuremath{{\mathrm{\,Me\kern -0.1em V}}}\xspace}
\newcommand{\gev}{\ensuremath{{\mathrm{\,Ge\kern -0.1em V}}}\xspace}
\newcommand{\tev}{\ensuremath{{\mathrm{\,Te\kern -0.1em V}}}\xspace}
\newcommand{\fm}{\ensuremath{{\mathrm{\,fm}}}\xspace}

\hypersetup{ 
    pdfnewwindow=true,
    colorlinks=true,
    linkcolor=royalblue,
    citecolor=royalblue,
    filecolor=royalblue,
    urlcolor=royalblue 
} 

\usepackage{fancyhdr}
\fancypagestyle{firstpage}{%
	
	\lhead{}
	\rhead{\small \begin{tabular}[t]{r}
        JLAB-THY-26-4756
	\end{tabular}}
}

\journal{Physics Letters B}
\crefformat{appendix}{#2#1#3}

\begin{document}
\begin{frontmatter}
\title{Finite-volume analysis of the \HD including left-hand-cut effects}
\author[odu,jlab]{Arkaitz~Rodas\,\orcidlink{0000-0003-2702-5286}}
\ead{arodasbi@odu.edu}
\author[odu,jlab]{Lin~Qiu\,\orcidlink{0000-0002-2683-8851}}
\ead{lqiu@odu.edu}
\author[uned]{C\'esar~Fern\'andez-Ram\'irez\,\orcidlink{0000-0001-8979-5660}}
\author[ub]{Vincent~Mathieu\,\orcidlink{0000-0003-4955-3311}}
\author[ub]{Gl\`oria~Monta\~na\,\orcidlink{0000-0001-8093-6682}} 
\author[messina,catania]{Alessandro~Pilloni\,\orcidlink{0000-0003-4257-0928}}
\author[jlab,ceem,indiana]{Adam~P.~Szczepaniak\,\orcidlink{0000-0002-4156-5492}}


\author{\\[.4cm] (Joint Physics Analysis Center)}

\affiliation[odu]{organization={Department of Physics, Old Dominion University}, city={Norfolk}, state={Virginia}, postcode={23529}, country={USA}}
\affiliation[jlab]{organization={Theory Center, Thomas  Jefferson  National  Accelerator  Facility}, city={Newport  News}, state={Virginia}, postcode={23606}, country={USA}}
\affiliation[uned]{organization={Departamento de F\'isica Interdisciplinar, Universidad Nacional de Educaci\'on a Distancia (UNED)}, postcode={E-28040}, city={Madrid}, country={Spain}}
\affiliation[ub]{organization={Departament de F\'isica Qu\`antica i Astrof\'isica and Institut de Ci\`encies del Cosmos, Universitat de Barcelona}, postcode={E-08028}, city={Barcelona}, country={Spain}}
\affiliation[messina]{organization={Dipartimento di Scienze Matematiche e Informatiche, Scienze Fisiche e Scienze della Terra, Universit\`a degli Studi di Messina}, postcode={I-98166}, city={Messina}, country={Italy}}
\affiliation[catania]{organization={INFN Sezione di Catania}, postcode={I-95123}, city={Catania}, country={Italy}}
\affiliation[ceem]{organization={Center for Exploration of Energy and Matter, Indiana University}, city={Bloomington}, state={Indiana}, postcode={47403}, country={USA}}
\affiliation[indiana]{organization={Department of Physics, Indiana University}, city={Bloomington}, state={Indiana}, postcode={47405}, country={USA}}
\begin{abstract}
We implement the finite-volume $N/D$ representation to study two-baryon interactions from lattice QCD data. We include the left-hand cut induced by one-pion exchange in this formalism, and study the \HD at the SU(3)$_\text{F}$-symmetric point, with a pion mass around $417\mev$. The $N/D$ formalism is then compared to the L\"uscher quantization condition, used to describe the same system via effective-range expansions. The results show a mild but statistically significant effect produced by the inclusion of the left-hand cut, especially on the binding energy of the \HD. 
\end{abstract}
\end{frontmatter}
\thispagestyle{firstpage}

\section{Introduction}
\label{sec:intro}
Understanding the low-energy behavior of quantum chromodynamics (QCD), and the hadron spectrum, including states that go beyond the conventional picture of hadrons as simple quark-antiquark pairs for mesons or three-quark combinations for baryons, remains a central goal of particle and nuclear physics. In recent years, a wealth of candidates for such exotic hadrons has been reported experimentally. These include near-threshold charmonium-like states such as the $X(3872)$~\cite{Belle:2003nnu,BaBar:2004oro}, the doubly charmed tetraquark candidate $T^+_{cc}(3875)$~\cite{LHCb:2021vvq,LHCb:2021auc}, hidden-charm pentaquark candidates $P_c$~\cite{LHCb:2015yax,LHCb:2019kea}, and light-quark candidates with spin-exotic quantum numbers such as $\pi_1(1600)$~\cite{JPAC:2018zyd,GlueX:2024erj}, $\eta_1(1855)$~\cite{BESIII:2022riz}, and the many light tetraquark and molecular candidates; see Refs.~\cite{Guo:2017jvc,Chen:2016qju,Chen:2022asf,Brambilla:2019esw,Pelaez:2025wma} for reviews. Many of these particles appear near thresholds, and can be affected by long-range forces. In fact, the one-pion exchange (OPE) mechanism was introduced as a compelling picture to describe some of these elusive states~\cite{Tornqvist:1991ks,Tornqvist:1993ng,Wang:2013kva,Karliner:2015ina,Baru:2015nea,PavonValderrama:2019nbk,Du:2023hlu}. 

Lattice QCD offers a path to study such systems from first principles. The finite-volume (FV) spectrum computed on the lattice can be related to infinite-volume (IFV) scattering amplitudes using quantization conditions, most prominently those originating from L\"uscher's formalism~\cite{Luscher:1986pf,Luscher:1990ux}; see Ref.~\cite{Briceno:2017max} for a review. However, while these methods have been broadly generalized~\cite{Rummukainen:1995vs,Kim:2005gf,Leskovec:2012gb,Briceno:2012yi,Briceno:2014oea}, standard implementations do not explicitly include nearby left-hand cuts, which may be important for accurate studies of states appearing near the branch points induced by one-particle exchange. This has motivated several recent developments aimed at incorporating left-hand-cut effects directly in FV analyses, including effective-field-theory constructions using plane-wave bases~\cite{Meng:2021uhz,Meng:2023bmz}, modified quantization conditions~\cite{Raposo:2023oru,Bubna:2024izx,Bubna:2025gsd,Raposo:2025dkb}, and treatments connected to three-body dynamics~\cite{Hansen:2024ffk,Dawid:2024dgy}. 

In this Letter, we present the first application of the FV $N/D$ formalism with a left-hand cut~\cite{Dawid:2024oey} to lattice-QCD two-body spectra. Building on the classic $N/D$ construction~\cite{Chew:1960iv,Bjorken:1960zz}, this representation is based purely on on-shell amplitudes, separates left-hand singularities into the numerator $N(s)$ and the unitarity cut into the denominator $D(s)$, and introduces an analytic representation based on dispersion relations. This makes this framework ideal for systems where long-range exchanges contribute to non-trivial dynamics.

The \HD is a long-standing exotic candidate, first proposed by Jaffe as a six-quark bound state~\cite{Jaffe:1976yi}. A deeply bound, compact \HD-like sexaquark has also been discussed as a possible dark-matter candidate within QCD~\cite{Gross:2018ivp,Farrar:2020zeo}. If located near the two-baryon threshold, the \HD would strongly affect low-energy baryon--baryon scattering and the spectrum of hypernuclei. Despite decades of experimental effort, no unambiguous signal has been established, with constraints coming from doubly-strange hypernuclei~\cite{Takahashi:2001nm,KEKE176:2009jzw,E373KEK-PS:2013dfg}, heavy-quarkonium decays~\cite{Belle:2013sba}, and femtoscopic correlation measurements in high-energy collisions~\cite{STAR:2014dcy,ALICE:2019eol}. Its existence and binding energy have also been studied in a wide range of phenomenological and effective-field-theory approaches, with conclusions that depend sensitively on the assumed dynamics and quark masses~\cite{Balachandran:1985fb,Yost:1985mj,Straub:1988mz,Kodama:1994np,Nakamoto:1997gh,Shen:2000qs,Haidenbauer:2011ah,Haidenbauer:2011za,Marietti:2022jil}. 

Lattice QCD has also studied this channel extensively in several quark-mass setups. In SU(3)$_\text{F}$-symmetric QCD, \ie, with \mbox{$m_u=m_d=m_s$}, early calculations reported a bound state~\cite{NPLQCD:2010ocs,Inoue:2010hs,Inoue:2010es}, while later studies showed a strong dependence on the lattice setup and continuum extrapolation~\cite{Hanlon:2018yfv,Green:2021qol}. Near the physical point, \mbox{HAL QCD Collaboration} found weakly attractive baryon-baryon interactions that do not support a bound or resonant \HD~\cite{HALQCD:2019wsz}. However, while this state appears near left-hand cut branch points, no analysis has yet been performed to assess the impact of the left-hand cuts on the extracted binding energy and other properties of the state.

The \HD system is especially suited to test a FV formalism with explicit left-hand cuts. The state lies close to the two-baryon threshold, and precise lattice spectra are available across several volumes and lattice spacings~\cite{Green:2021qol}. We study the SU(3)$_\text{F}$-symmetric, flavor-singlet two-baryon channel, with \mbox{$m_\pi\approx 417\mev$}, modeling the left-hand cut through one-meson exchange.\footnote{Under exact $\mathrm{SU}(3)_{\mathrm{F}}$ symmetry, all members of the pseudoscalar-meson octet are mass degenerate with the pion. For example, for the $\Lambda$--$\Lambda$ interaction the exchanged meson is the $\eta_8$. Throughout the text, we refer to this degenerate one-meson-exchange contribution as pion exchange.} We compare the resulting $N/D$ description of the lattice spectra with ordinary L\"uscher analyses based on effective-range expansion (ERE) parameterizations.

In the following, we present in~\cref{sec:fv_formalism} the $N/D$ and L\"uscher formalisms, the lattice data, and the procedure used to fit the data. In~\cref{sec:params} we present our infinite-volume parameterizations. Finally, we present our fit results, extract low-energy parameters, and \HD pole properties in~\cref{sec:results}.

\section{Finite-volume approach}
\label{sec:fv_formalism}
We first introduce the ordinary L\"uscher quantization condition~\cite{Luscher:1990ux,Rummukainen:1995vs,Kim:2005gf,Leskovec:2012gb,Briceno:2012yi,Briceno:2014oea,Briceno:2015tza} where all left-hand-cut effects are ignored, and then compare with the $N/D$ quantization condition introduced in Ref.~\cite{Dawid:2024oey}, which allows for the explicit implementation of left-hand cuts. 

The L\"uscher quantization condition can be expressed as
\begin{equation}
\label{eq:luscher}
K^{-1}(E)+F( E, \boldsymbol{P}, L)=0,
\end{equation}
where $E$ and $\boldsymbol{P}$ are the energy and momentum in a cubic, periodic volume of box size $L$. $K^{-1}$ is the inverse of an IFV two-body $K$-matrix~\cite{Aitchison:1972ay}, and $F$ represents a FV function of the lattice, given by
\begin{align}
\label{eq:Fpv}
&F(E, \boldsymbol{P}, L)=\left[\frac{1}{L^3} \sum_{ \boldsymbol{k} } -\text {p.v.} \int \frac{d \boldsymbol{k} }{(2 \pi)^3}\right] \nonumber \\
&\times \frac{16 \pi}{2 \omega_1( \boldsymbol{k} ) 2 \omega_2( \boldsymbol{P} - \boldsymbol{k} )\left(E-\omega_1( \boldsymbol{k} )-\omega_2( \boldsymbol{P} - \boldsymbol{k} )\right)},
\end{align}
where $\omega_i(\boldsymbol{k})=(\boldsymbol{k}^2+m_i^2)^{1/2}$ is the on-shell energy of the scattering particle of mass $m_i$ and momentum $\boldsymbol{k}$, and the principal-value prescription is used.\footnote{Note that we include an extra factor $32 \pi$ to account for our unitarity condition in the partial-wave amplitudes, and a factor $1/2$ associated with identical-particle scattering.} 
Above the left-hand cuts, the L\"uscher quantization condition is equivalent to the $N/D$ formalism, as shown in Section S3 of the Supplemental Material~\cite{SupplementalMaterial}. In the IFV $N/D$ method, the numerator $N(s)$ includes the left-hand cuts while the denominator $D(s)$ contains only the right-hand cuts given by unitarity in the $s$-channel. The numerator is only affected by exponentially suppressed effects when evaluating it in the FV. The $K^{-1}$ above can be substituted by a dispersive representation that relates it to the IFV $D$ function times the inverse of the numerator. In so doing, one can multiply~\cref{eq:luscher} by $N$, and manipulate $F$ algebraically to cancel the principal value integral with the corresponding dispersion from $K^{-1}$. The resulting sum corresponds to the finite-volume $D_{\text{FV}}$ piece, and the quantization condition reads
\begin{equation}
D_{\text{FV}}\left(s, \boldsymbol{P} \right)=0.
\end{equation}

The denominator is free from the singularities in the left-hand cut and can be analytically continued in that energy region, providing the sought condition. We use a once-subtracted dispersive representation, subtracted at \mbox{$s=0$}, for the denominator
\begin{align}
\label{eq:Dfv}
D_{\text{FV}}\left(s, \boldsymbol{P} \right)&= 1 + p_n(s) \nonumber \\
&-\frac{16 \pi}{L^3} \sum_{ \boldsymbol{k} } L ( \boldsymbol{k} , \boldsymbol{P} ) \frac{s}{E^{\star}( \boldsymbol{k} )^2}\frac{ N \left(\boldsymbol{k}^{\star}, \boldsymbol{P} \right)}{E^{\star}( \boldsymbol{k} )^2-s} ,
\end{align}
where $p_n(s)$ can be an $n$th-degree ($n\ge 1$) polynomial in $s$, starting at first order, with a possible additional sum of Castillejo-Dalitz-Dyson (CDD) poles~\cite{Castillejo:1955ed}.

We have normalized the denominator to one at $s=0$ and $L(\boldsymbol{k},\boldsymbol{P})$ is derived as\footnote{In this Letter, we label all quantities boosted into the center-of-mass frame with a superscript "$\star$".}
\begin{equation}
L ( \boldsymbol{k} , \boldsymbol{P} )=\frac{2 E( \boldsymbol{k} )}{2 \omega_1( \boldsymbol{k} ) 2 \omega_2( \boldsymbol{P} - \boldsymbol{k} )},
\end{equation}
where $E(\boldsymbol{k})=\omega_1(\boldsymbol{k})+\omega_2(\boldsymbol{P}-\boldsymbol{k})=({E^{\star}}^2+\boldsymbol{P}^2)^{1/2}$ is the total scattering energy with $E^{\star}(\boldsymbol{k})=\omega_1(\boldsymbol{k}^{\star})+\omega_2(\boldsymbol{k}^{\star})$.

The order of $p_n(s)$ encodes subtraction freedom in the dispersive representation, and possible CDD-pole contributions can also be included. The explicit choice used in this work is explained in~\cref{sec:params}.

Our analysis is based on energy levels determined from correlation functions in Ref.~\cite{Green:2021qol}, using eight gauge ensembles from the Coordinated Lattice Simulations (CLS) $N_f=2+1$ program~\cite{Bruno:2014jqa,Bruno:2016plf}. The FV spectra were extracted from correlation matrices built using the distillation technique~\cite{HadronSpectrum:2009krc} and a variational generalized eigenvalue problem (GEVP) analysis~\cite{Luscher:1990ck,Blossier:2009kd}, in the rest frame and several moving frames. The pion and baryon masses for each ensemble are provided in the supplemental material of Ref.~\cite{Green:2021qol}. These masses vary by only a few percent. In our analysis, the pion mass is fixed to the average value over all the ensembles. For the baryon mass, however, we first use the ensemble masses when transforming the input energy levels to momentum, and then fix the value to the average over all ensembles used when performing the finite-volume fits. We only study the effect of varying the baryon mass as a systematic effect on our low-energy parameters and pole positions for the \HD.

\begin{figure*}
\centering
\includegraphics[width=0.9\textwidth]{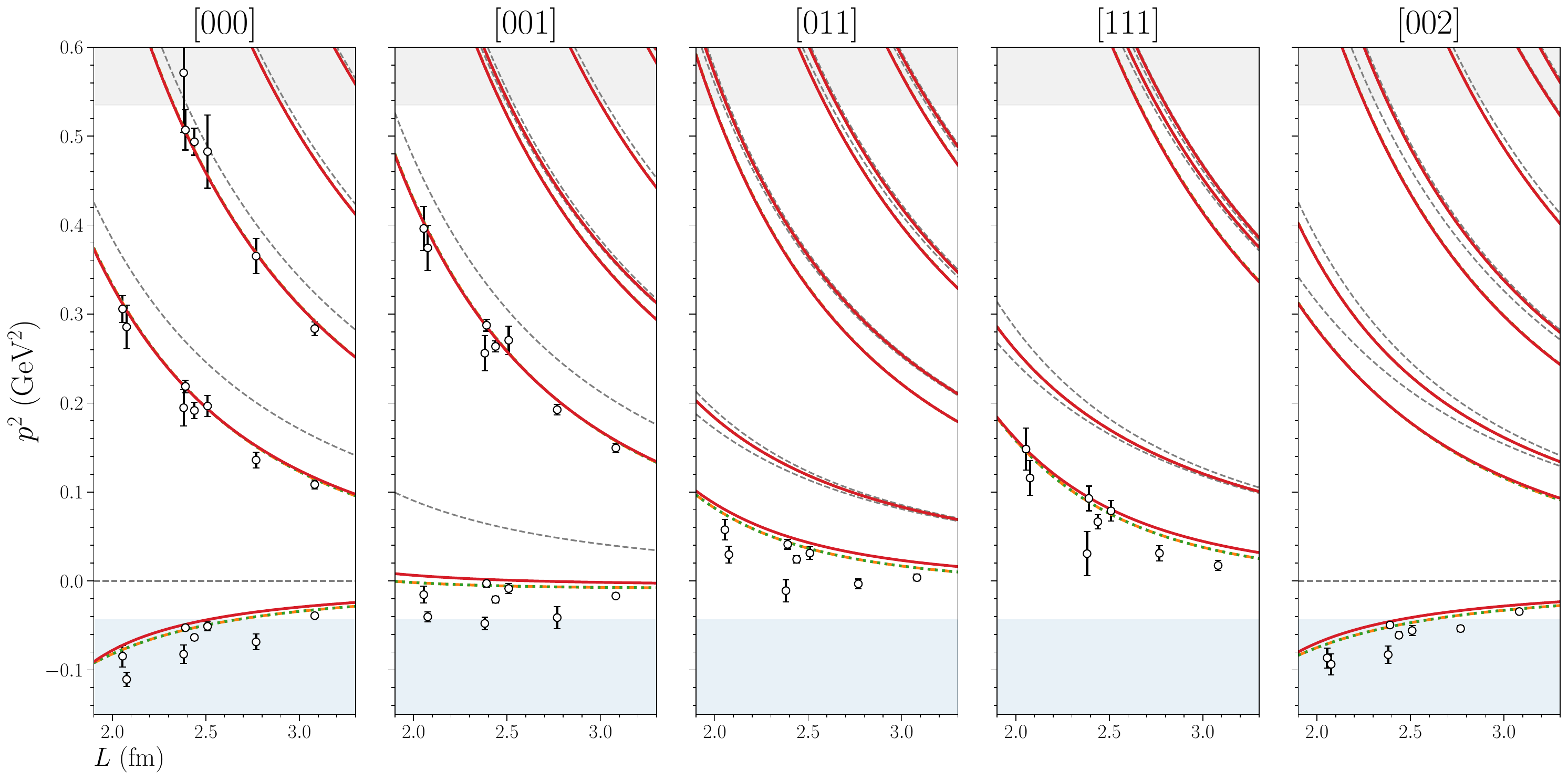} 
\caption{
Comparison of the spectra obtained by our three different models, fitted to the FV lattice spectrum: ERE (orange dashed), ERE with a Chew-Mandelstam term (green dotted) and $N/D$ (red solid). No model uncertainties are shown for clarity of presentation. Grey dashed lines display the non-interacting energy levels. The grey shaded area represents the inelastic region where three-body effects may become relevant~\cite{Briceno:2018aml}, and the blue area represents the region where the OPE left-hand cut is located.}
\label{fig:spectrum_a}
\end{figure*} 

\subsection{Fitting procedure}
\label{subsec:fv_fit}
We use both the single- and two-baryon energy levels given in Ref.~\cite{Green:2021qol}. The available data contain $10^3$ bootstrap samples for each energy level. Two solutions for each level are available. The preferred solution is used to establish central values and statistical errors. The alternative solution is used to establish a systematic uncertainty.

We fit all models to the same set of 61 data points between the lowest energy level and the three-body inelastic threshold, given by \mbox{$E^{\star}=2 m_B+m_\pi$}, including levels in the left-hand-cut region. As these levels come from 8 different lattice ensembles, small systematic errors exist when determining the single-hadron masses. In the following, we fix the pion mass to the average over the lattice ensembles, which gives \mbox{$m_\pi\approx 417\mev$}. We do not perform any systematic study of the pion mass dependence, as we observe mild left-hand-cut contributions to the \HD.

While the original levels were presented in the energy variable, we follow the suggestion in Ref.~\cite{Green:2021qol} and calculate their $p^2$ value instead, \ie, 
\begin{equation}
p^2={E^{\star}}^2\slash 4 - m_B^2,
\end{equation}
where for each ensemble we use the corresponding $m_B$ value to remove associated systematic effects. With the effect of $m_B$ variations removed, we can study the FV behavior using the averaged baryon mass \mbox{$m_B \approx 1.18\gev$} with the exception of the \HD pole-related parameters, where we perform further fits using different baryon masses, as this state appears close to threshold.

The fits are obtained by minimizing
\begin{equation}
\chi^2 = \sum_{i, j}\left(p_i^2-p_{\text{FV}, i}^2\right) \Sigma_{i j}^{-1}\left(p_j^2-p_{\text{FV}, j}^2\right),
\end{equation}
where $p^2_{\text{FV}}$ is the value predicted by the corresponding quantization condition in each proposed model; $\Sigma_{i j}$ is the data covariance with $i,j$ the indices of all $61$ energy levels in the fits.
The data covariance includes both statistical and systematic parts, \ie, \mbox{$\Sigma_{i j}=\Sigma_{\text{stat},\, i j}+\Sigma_{\text{syst},\, i j}$}. The statistical piece is obtained directly from the bootstrap samples for the energy levels and is therefore a block-diagonal matrix, with each block corresponding to one of the ensembles. The systematic covariance matrix reads
\begin{equation}
\Sigma_{\text{syst},\, i j}=\left(\delta p^2\right)_i\left(\delta p^2\right)_j,
\end{equation}
where indices $i$ and $j$ are restricted to energy levels belonging to the same ensemble, and thus the matrix is also block diagonal and can be added directly to the statistical one. We find that the systematic errors are not large and the $\chi^2$ values vary only mildly if they are ignored. Extra levels exist in this system, but they either correspond to a different spin system, or might be affected by higher partial waves, which we do not include in this work.

We note that each lattice ensemble is characterized by a different spacing $a$. In this sense, the energy levels are not determined in the continuum, and the FV formalisms presented above are not rigorous. We adopt the phenomenological prescription provided in Ref.~\cite{Green:2021qol} to account for lattice spacing effects. For a given parameterization, we consider each parameter $\xi_i$ as a function of $a$, and expand to first order in $a^2$ as\footnote{This is the expected leading-order spacing correction for improved lattices like the ones studied here.}
\begin{equation}
\xi_i=\xi_{i0}+\xi_{ia}a^2.
\label{eq:discrete_pars}
\end{equation}
In this approach, $\xi_{ia}$ gives a rough idea of how relevant discretization effects are. It also doubles the number of free parameters we use for our description of the finite-volume energy levels, and we therefore expect to see substantial improvements in fits when following this procedure.

It is worth mentioning that this prescription is not a rigorous treatment of finite-spacing effects. While some work exists in the literature studying discretization effects~\cite{Hansen:2024cai}, no complete, algebraic solution is known to us. For that reason, we checked these effects with an alternative approach. Instead of introducing explicit \mbox{$a$-dependent} parameters in the FV formalisms, one may treat the energy levels on the lattice as \mbox{$a$-dependent} ones and then extrapolate them directly into the continuum, level by level. Operationally, this requires grouping together the same FV levels across the different lattice volumes and spacings, and using those correlated determinations to extract the corresponding continuum values, given a model.
The resulting continuum-extrapolated dataset can be analyzed with no explicit $a$ dependence in the amplitudes. We find our $N/D$ model results to be stable compared with the $a$-dependent procedure explained above, while those from our ERE models show larger deviations.

Finally, the central values and statistical errors of our parameters come directly from \texttt{iminuit}~\cite{James:1975dr,iminuit}. While the lattice QCD data are provided via bootstrapping, we are dealing with uncorrelated ensembles, and thus the covariance matrix is block diagonal, which justifies our fitting strategy. In what follows, uncertainties propagated from the fitted parameters are referred to as statistical uncertainties, although the fits themselves use the full data covariance matrix described above. All fits reported in this work satisfy the convergence criteria of the minimization algorithm.

\section{Parameterizations}
\label{sec:params}
We assume that the two-baryon system can be described via elastic scattering above threshold, within a single partial wave $S$. In this approximation, the single partial wave reads
\begin{equation}
t_0(s)= \frac{1}{\rho(s)\cot \delta_0(s) - i\rho(s)}, \label{eq:swave}
\end{equation}
where $\delta_0(s)$ denotes the phase shift of the $S$ wave, and $\rho(s)$ is the customary phase space, given, in this case, by
\begin{equation}
\rho(s)\equiv \frac{2 p(s)}{\sqrt{s}}= \sqrt{\frac{s-s_{\text{thr}}}{s}}, \label{eq:phasespace}
\end{equation}
with \mbox{$s_{\text{thr}}\equiv 4 m_B^2$} the two-baryon threshold.\footnote{We note that this phase-space convention requires an extra $32 \pi$ factor in the quantization conditions above, in contrast to the amplitudes defined in Refs.~\cite{Dawid:2024oey,Kim:2005gf}.} 

For the analytical continuation of the amplitude, especially below threshold, we use three different parameterizations. The first two neglect the existence of left-hand cuts. In this case, $[K(s)]^{-1}$ can be expanded in powers of the $p^2$ variable. We refer to these parameterizations as EREs, although our fits retain terms through $p^4$, beyond the customary two-parameter effective-range expansion. Explicitly, we use
\begin{align}
\rho(s)\cot \delta_0(s) &  = \frac{2}{\sqrt{s}} \left[\frac{1}{c_0} + c_1 \, p(s)^2 + c_2 \, p(s)^4 \right],
\label{eq:pcotdelta}
\end{align}
where $c_j$ are the expansion coefficients to be fitted, which, unless stated otherwise, follow the prescription given by~\cref{eq:discrete_pars}. These ERE parameterizations apply to the lattice QCD data via the ordinary L\"uscher formalism, for which we implement $\left[K(s)\right]^{-1}$ in~\cref{eq:luscher}. 

For the first ERE form, we keep the naive phase space from~\cref{eq:phasespace}, which leads to $\left[K(s)\right]^{-1}=\rho(s)\cot \delta_0(s)$. 
This phase space introduces an artificial pole at \mbox{$s=0$} by construction. An alternative dispersive representation, which provides the same imaginary part as required by elastic unitarity but avoids this issue, is given by the Chew-Mandelstam term~\cite{Chew:1960iv}, subtracted at threshold here,
\begin{equation}
I(s)= -\frac{s-s_{\text{thr}}}{\pi} \int_{s_{\text{thr}}}^{\infty} d s^{\prime} \frac{\rho\left(s^{\prime}\right)}{\left(s^{\prime}-s\right)\left(s^{\prime}-s_{\text{thr}}\right)},
\end{equation}
whose analytic expression can be found in Refs.~\cite{Oller:1998zr,Wilson:2014cna}. 

Our second ERE model uses this dispersive phase space $I(s)$, replacing $-i \rho(s)$. In this case, $\left[K(s)\right]^{-1}$ includes an extra $\text{Re}\, I(s)$ function. Due to the real part of the Chew-Mandelstam term, the parameters in this expansion are modified with respect to the previous case.

When there is a left-hand singularity, the above expansions converge only in a region centered at threshold, within a radius smaller than the distance to the left branch point, which for the OPE produced in this baryon-baryon channel, is rather small. Thus, it is not adequate to describe data far from threshold with these ERE parameterizations. In fact, recent work has emphasized the impact of left-hand singularities on effective parameterizations, and the necessity to include such effects to obtain robust pole extractions from lattice spectra; see, e.g., Refs.~\cite{Du:2023hlu,Du:2024gzw,Wang:2026ups,Liu:2026xrk}.

Additionally, the L\"uscher formalism cannot be applied directly for energy levels that appear near or on top of left-hand cuts. Thus, we adopt the $N/D$ framework as the third amplitude model~\cite{Chew:1960iv,Bjorken:1960zz}, \ie, we parameterize
\begin{equation}
\label{eq:noverd}
t_0(s)=\frac{N(s)}{D(s)},
\end{equation}
which incorporates both the right-hand cut in the denominator $D(s)$ and the left-hand cut in the numerator $N(s)$, produced by the partial-wave projection of a one-particle crossed-channel exchange.

To obtain a comprehensive solution, one usually solves a set of integral equations with respect to both $N(s)$ and $D(s)$, given the discontinuity along the left-hand cut and subtractions~\cite{Oller:1998zr,Oller:2018zts}. Instead, following the generalization in Ref.~\cite{Du:2024gzw}, we tested a variety of parameterizations and finally adopted the numerator as a constant along with a logarithmic term accounting for the OPE mechanism in an $S$ wave, arising from one-pion exchange in the $t$- or $u$-channels. This choice allows us to introduce an $N/D$ parameterization that reverts to the ERE with Chew-Mandelstam term presented above, when the left-hand cut is neglected. Explicitly, we include a logarithmic term that results from projecting the OPE potential from the crossed channel into the $s$-channel as~\footnote{In principle, $L_t$ and $L_u$ are distinguished, but for equal-mass scattering, both produce the same branch point.}
\begin{align}
L_t\left(s\right) &\equiv \frac{1}{2} \int_{-1}^{+1} \frac{d\! \cos \theta}{t-m_\pi^2} 
=-\frac{1}{s-s_{\text{thr}}} \log \frac{s-s_{\text{thr}}+m_\pi^2}{m_\pi^2}. \label{eq:Lt}
\end{align}
As a result, 
\begin{equation} 
N(s)=n_0 + g_0^2\, L_t(s).
\label{eq:onceN}
\end{equation}
Given that we use a once-subtracted dispersion relation for $D(s)$, $N(s)$ must asymptote to at most a constant at infinity. Using higher orders for $p_n(s)$ thus forces the partial wave to approach zero as $s$ increases. In this work, we implement ERE parameterizations up to order $p^4$, and an $N/D$ representation that shares the same asymptotic behavior, as required by the lattice QCD data.

Since our numerator asymptotes to a constant, the denominator is a once-subtracted dispersive result, where we subtract at $s=0$ for convenience. We also consider higher-order terms with real coefficients in $p_n(s)$, to add flexibility to the fits and to respect the same asymptotic behavior as the ERE forms, \ie,
\begin{equation}
D(s)=1 + d_0 s + d_1 s^2 +\frac{s}{\pi}\int_{s_{\text{thr}}}^\infty d s^\prime \, \frac{\rho(s^\prime)\, N(s^\prime)}{s^\prime \left(s-s^\prime\right)},
\label{eq:onceD}
\end{equation}
and thus we have four free parameters $n_0,\ g_0,\ d_0$ and $d_1$ in our $N/D$ model to be fitted in the continuum.

The amplitude defined via~\cref{eq:noverd,eq:Lt,eq:onceN,eq:onceD} satisfies elastic unitarity and falls to zero sufficiently fast as $s$ increases, which turns out to be needed to properly describe the lattice data, as seen below. We do not include any explicit CDD pole in this formalism, but one can check a posteriori that the resulting numerator does not cross zero above threshold, confirming the absence of any of these poles; see the Supplemental Material~\cite{SupplementalMaterial} for details. Note that for \mbox{$g_0^2=0$}, this formula reverts to the ERE form, where the dispersion relation produces a Chew-Mandelstam phase space. In this limit, the only difference between this model and the ERE ones is the $2/\sqrt{s}$ term appearing inside~\cref{eq:pcotdelta}. Therefore, by comparing this parameterization with the previous two, we obtain a direct understanding of whether the left-hand-cut effects are relevant and how they affect the amplitudes.

\begin{figure*}[!htb]
\centering
\includegraphics[width=0.8\textwidth]{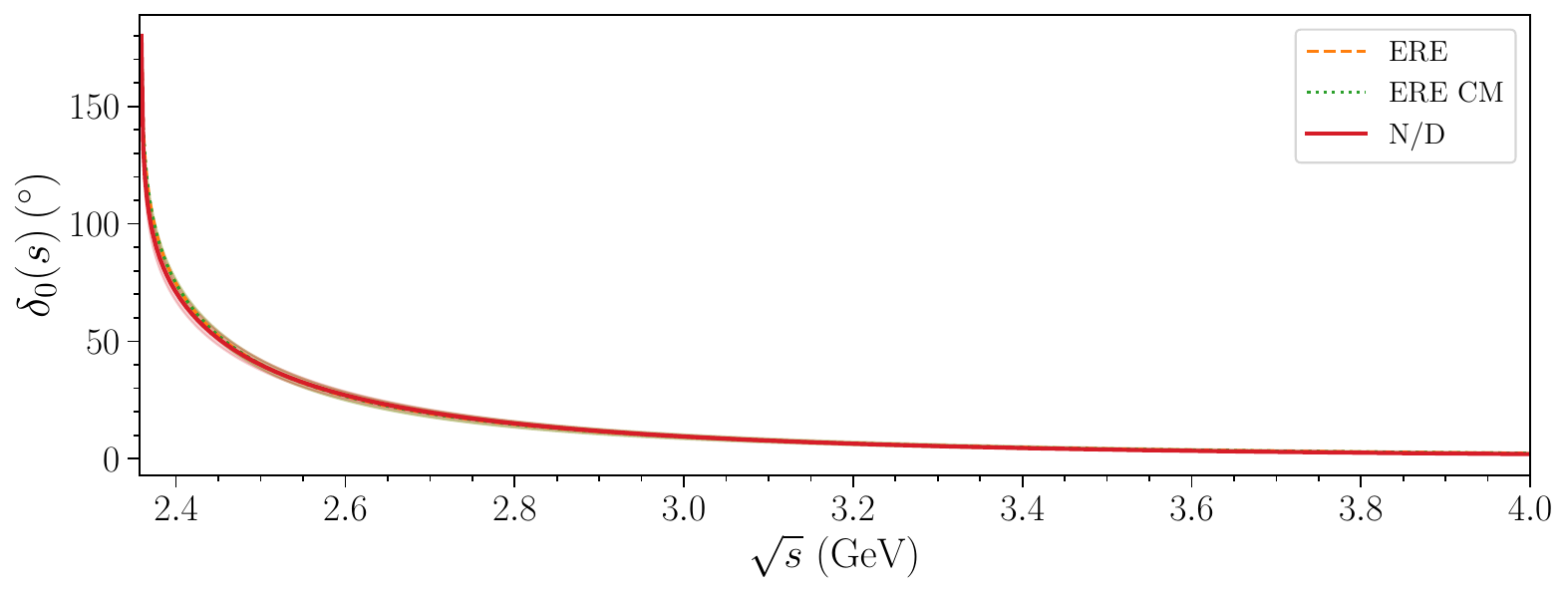}\\
\vspace{-0.2cm}
\includegraphics[width=0.8\textwidth]{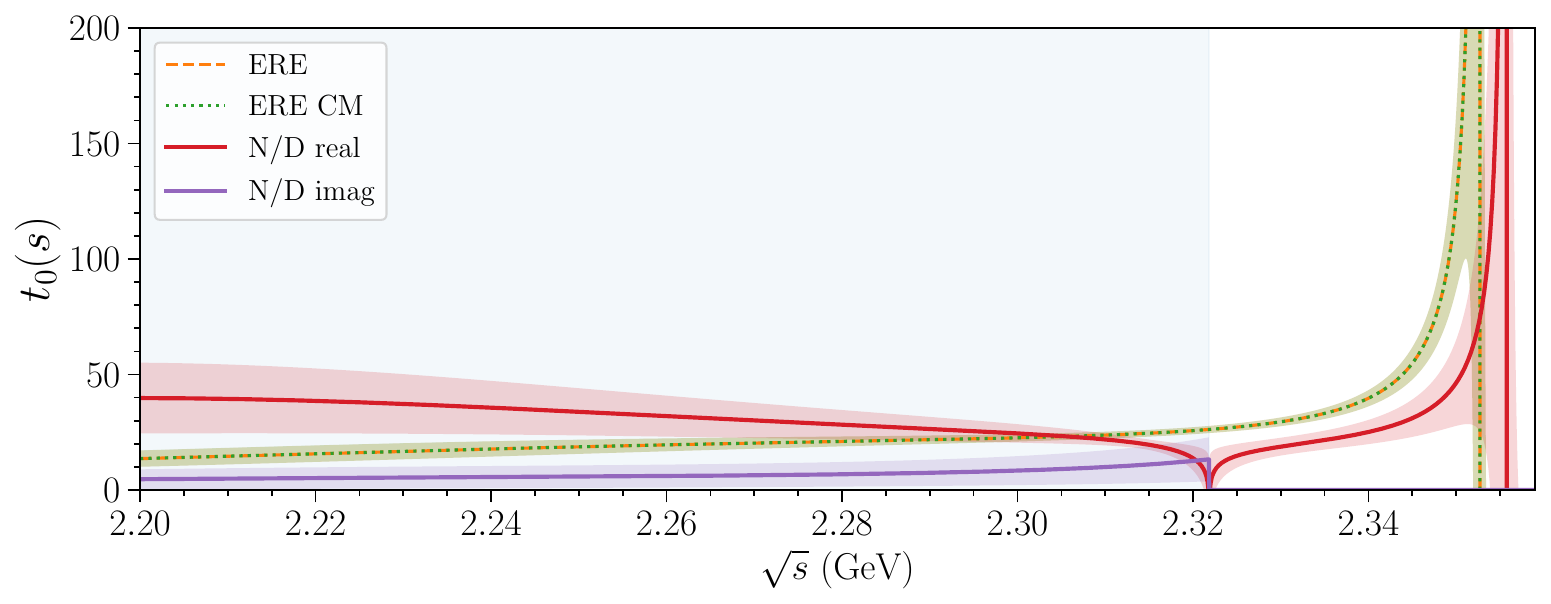}
\caption{ Comparison of the IFV amplitudes obtained from the three models fitted to the FV lattice spectrum. The figures include central values and statistical errors. Top: phase shifts in the physical region. Bottom: amplitudes below threshold. ERE denotes the effective-range expansion and ERE CM denotes the one with a Chew-Mandelstam term. The blue shaded area represents the region where the OPE left-hand cut is located. Note the behavior of the $N/D$ model in that region, in contrast with the ERE models, which do not contain any left-hand cut. All models feature a pole a few MeV below threshold.} 
\label{fig:amp_comparison}
\end{figure*}

\section{Results}
\label{sec:results}

First, we fit the data including $a$ dependence for the different models, but only show the results in the continuum, where all parameters at $\mathcal{O}(a^2)$ are set to zero when evaluating the scattering amplitudes. When uncertainties are included in the plots, these are the statistical uncertainties propagated from the fitted parameters, and do not include any systematic contribution. In~\cref{fig:spectrum_a}, we plot the predicted spectra for the three different models. For clarity, the uncertainties are not displayed. As the $a^2$ related parameters are turned off, the distance between the predicted energy levels and corresponding data points cannot be directly related to the resulting $\chi^2$, which does include $a$ dependence and covariance. Nonetheless, the fits still produce reasonable spectra and our models largely follow the data pattern.

As expected, the fits only differ in the lowest energy region, and are almost identical above it. Note that, for this system, interacting energy levels always have a corresponding non-interacting level in the vicinity and no extra levels are observed. Additionally, all low-energy levels exhibit attraction, which becomes milder as the energy increases. For the highest energies considered in this work, energy levels are compatible with non-interacting energies, hinting at an amplitude suppressed at higher energies. As can be seen in~\cref{fig:spectrum_a}, the ERE models are indistinguishable to the eye, while the $N/D$ model differs in the low-energy region. This can be explained by the presence of the OPE in the $N/D$ formalism. Interestingly, the $N/D$ model seems to slightly shift the lowest energy levels upwards with respect to ERE models. We also note that the ground-state levels exhibit significant volume dependence, \ie, the predicted $p^2$ is closer to zero when $L$ increases. As a result, if a bound state exists, it is expected to appear near threshold.

All three models provide an accurate description of the data with comparable \mbox{$\chi^2/N_{\text{dof}}\sim 0.76$} when the $a$ dependence is included. The ERE models contain three parameters in the continuum (six in total), while the $N/D$ model has four in the continuum (eight in total). The explicit fit results, including correlations, are presented in Section S1 of Ref.~\cite{SupplementalMaterial}.

Fits with fewer parameters in the continuum degrade substantially, as these parameterizations are not flexible enough to describe the data, especially those near the left-hand cut. Fits with more parameters do not produce significant improvements, some even worsen $\chi^2/N_{\text{dof}}$, and none are considered any further. Finally, some of the $\xi_{ia}$ parameters deviate significantly from zero, hinting at a sizable dependence. Therefore, we checked the data description when neglecting $a$ dependence, and still found good agreement with the data for all three main models, though the $\chi^2$ increases by about a factor of two. Further information on these fits can be found in Section S2 of Ref.~\cite{SupplementalMaterial}.

The resulting phase shifts are shown in the top panel of~\cref{fig:amp_comparison}. 
All of them are largely compatible within small uncertainties. All parameterizations feature a pole below threshold between the left- and right-hand cuts, which can be associated with a bound state and explains the rapid variation of the phase shift close to threshold. As the energy increases, the phase shifts decrease smoothly to values close to zero, characteristic of a weakly interacting energy region. These line shapes could not be well reproduced if we had used fewer parameters in the continuum limit. As a result, all models with fewer parameters provided much larger $\chi^2$ values, and did not describe the data accurately. For the $N/D$ model, one extra parameter is needed to account for the inclusion of the OPE, which barely modifies the phase shift in the physical region, but is relevant for the behavior of the amplitude below threshold, as we explain below.

In the bottom panel of~\cref{fig:amp_comparison}, we plot the amplitudes for the three models below threshold, including the region where the left-hand cut is located. The differences among the models are visible below threshold. First, all models feature a single pole, signaled by the divergent behavior of the amplitudes below threshold in the energy region of interest. As required by analyticity, the amplitudes are real in the energy region between the left- and right-hand cuts. In the left-hand-cut region, the amplitudes in ERE models are always real, while the $N/D$ model features an imaginary part starting at the OPE branch point. The appearance of an imaginary part is a consequence of the crossing-related intermediate states going on shell, e.g., the virtual pion. While the EREs are featureless, the $N/D$ contains richer physics in that area. In our fits, the OPE-related parameter $g_0$ is small, which explains the small $N/D$ imaginary part compared to its real part. However, the results below threshold still differ from those in ERE models.

\begin{figure*}[!bht]
\centering
\includegraphics[width=0.82\textwidth]{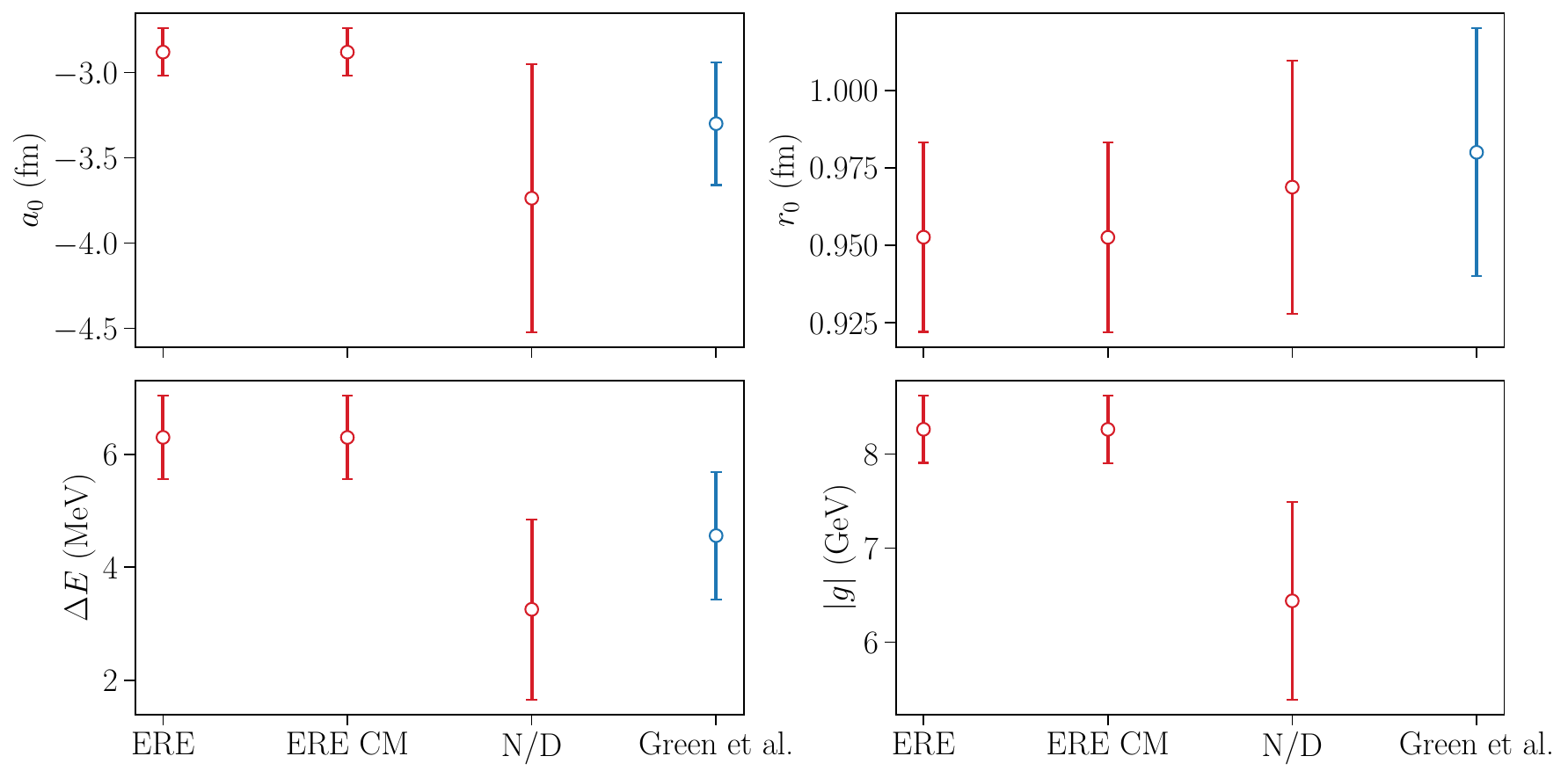}
\caption{Observables extracted from the three different models used to describe the FV lattice spectrum including spacing dependence (red), and the corresponding results cited from the near-threshold fitting in Ref.~\cite{Green:2021qol} (blue). Top: scattering length $a_0$ (left) and effective range $r_0$ (right). Bottom: binding energy $\Delta E$ of the \HD (left) and modulus of its coupling (right). The panels include central values and statistical errors.}
\label{fig:poles_a}
\end{figure*} 

\subsection{Low-energy parameters}
\label{subsec:lep}
The \HD is expected to appear in the vicinity of the two-baryon threshold, affecting the behavior of the partial wave in that region. The low-energy scattering parameters play a significant role in explaining the properties of states appearing in this region. Once our fits are performed, we can calculate and compare these parameters at threshold, which we can later relate to the properties of the \HD.

To address the near-threshold behavior, the standard practice is to calculate the scattering length $a_0$ and the effective range $r_0$, given, in our convention, by
\begin{subequations}
\label{eq:lec}
\begin{align}
a_0&=\left. \frac{2}{\sqrt{s}}\,\frac{1}{t_0^{-1}(s)+i\rho(s)}\right|_{s=s_{\text{thr}}},\\
r_0 &= 4\frac{d}{ds}\left[\sqrt{s}\,\left(t_0^{-1}(s)+i\rho(s)\right)\right]_{s=s_{\text{thr}}}.
\end{align}
\end{subequations}

For the $N/D$ model in the continuum, the low-energy parameters are evaluated to be
\begin{subequations}
\begin{align}
a_0&=\left(-3.737 \pm 0.787 \pm 0.001\right)\fm,\\
r_0&=\left(0.969 \pm 0.041 \pm 0.001\right) \fm,
\end{align}
\end{subequations}
where the central values are obtained based on the average baryon mass $m_B$ over all the lattice ensembles, the first uncertainties are the statistical uncertainties propagated directly from the fitted parameters, and the second ones account for the systematic variation associated with the averaged baryon mass within $1\sigma$.
We note that the parameters are qualitatively similar to the physical deuteron case, also a two-baryon system; in particular, the effective range is positive.

In the top panel of~\cref{fig:poles_a}, we present the calculated $a_0$ and $r_0$ from the three models fitted including spacing dependence, along with the corresponding results fitted within the near-threshold region ($|p^2|\lesssim m_{\pi}^2/4$) in Ref.~\cite{Green:2021qol} for comparison. The error bars correspond to just the statistical uncertainties propagated from the fitted parameters. We note that the results in our ERE models deviate slightly from the results in Ref.~\cite{Green:2021qol}, by around $13\%$ for $a_0$ and $3\%$ for $r_0$ in central values, considering that only two parameters were used therein. 

The effective range in our N/D model is compatible with all other results, while $a_0$ deviates above $10\%$ with respect to Ref.~\cite{Green:2021qol}, but with larger uncertainty. A negative $a_0$ implies a possible bound \HD, whose binding momentum can be estimated by $\mathcal{O}(-1/a)$ in a purely molecular scenario~\cite{Guo:2017jvc,Baru:2021ldu}, and thus suggests its binding energy to be only a few \mev. Compared to our ERE models, both $a_0$ and $r_0$ are larger in absolute value in our $N/D$ model, which hints at a smaller binding energy and a larger molecular component.

\subsection{\HD properties}
\label{subsec:poles}
All of our fitted parameterizations feature a shallow bound state below the two-particle threshold. Near the pole, the partial wave behaves as
\begin{equation}
t_0(s)\approx -\frac{r_p}{s-s_p},
\end{equation}
where $r_p$ is the residue of the amplitude at the pole position $s_p$, which must be a positive real number for a physical bound state. We relate this residue to the modulus of the physical coupling by $|g|=\sqrt{16\,\pi \,r_p}$~\cite{Garcia-Martin:2011nna, Pelaez:2020gnd}.

For the $N/D$ model in the continuum, our pole results are
\begin{subequations}
\begin{align}    
\Delta E&= (3.26 \pm 1.60 \pm 0.04)\,\mev, \\
\quad |g|&=  (6.44 \pm 1.05 \pm 0.04) \,\gev, 
\end{align}
\end{subequations}
with $\Delta E\equiv 2m_B-\sqrt{s_p}$ the binding energy of the \HD. The central values and the statistical and systematic uncertainties are obtained in the same way as for $a_0$ and $r_0$. We note that our binding energy is compatible with the bound-state energy inferred from $\Lambda\Lambda$ femtoscopy correlations~\cite{ALICE:2019eol}.

In the bottom panel of~\cref{fig:poles_a}, we compare the binding energies and couplings calculated from the models, where the error bars again represent only the statistical uncertainties given that the systematic ones are negligible, along with the binding energy cited from Ref.~\cite{Green:2021qol}. Similarly, our fits performed using ERE models are almost identical to each other and thus the Chew-Mandelstam modification is mostly irrelevant. More importantly than the low-energy parameters, the pole results in ERE models deviate largely from that in Ref.~\cite{Green:2021qol}, e.g., the binding energies here are larger than $6\mev$ in central values while the cited one is only \mbox{$4.56\pm1.13\pm0.63\mev$}.

The results in the $N/D$ model are incompatible with the other two in our work, both in the pole position and coupling. Its slightly smaller binding energy agrees again with the cited one, showing partial consistency between the energy-restricted ERE models and the $N/D$ one, when fitted to the proper energy levels individually. Given the tension between our ERE and $N/D$ parameterizations, we must conclude that the left-hand cut, incorporated using the $N/D$ method as an OPE, has a limited but non-negligible effect on the extraction of these observables from FV data.

For a two-body bound system, one can estimate the compact/molecular component by studying its so-called compositeness from the above observables. Weinberg's compositeness criterion~\cite{Weinberg:1962hj,Weinberg:1965zz} was established to study the deuteron. It is only justified by evaluating either the binding energy and physical coupling, or the scattering length and effective range, when the ERE applies and is truncated at $\mathcal{O}(p^2)$. Later, the above criterion was generalized in a series of works~\cite{Baru:2003qq,Matuschek:2020gqe,Li:2021cue,Esposito:2021vhu,Albaladejo:2022sux} to account for different circumstances including bound, virtual, and resonant states. Since we use three parameters in the continuum in the ERE models, as well as the contribution of the left-hand cut, we refrain from using Weinberg's compositeness criterion $Z$, which only applies to bound states far away from any branch point but the right-hand cut. Instead, we adopt the criterion $\bar{X}_A=1-\bar{Z}_A=1/\sqrt{{1+|2r_0/a_0|}}$ from Ref.~\cite{Matuschek:2020gqe} as a qualitative source of information on the nature of composite systems. Using $a_0$ and $r_0$ in the continuum from the $N/D$ model, we obtain that $\bar{X}_A\sim 0.81$, with small errors, while a slightly smaller $\bar{X}_A \sim 0.78$ from the ERE parameterizations. Finally, we note that these values are similar to those obtained for the physical deuteron.

\section{Summary}
\label{sec:summary}
We have presented the first application of the $N/D$ finite-volume formalism to lattice QCD data. We used a large collection of lattice QCD energy levels across different ensembles from Ref.~\cite{Green:2021qol} to study whether the inclusion of the left-hand-cut effects via one-pion exchange is relevant for determining the pole position and coupling of the \HD at the SU(3)$_\text{F}$-symmetric point. We compared the $N/D$ method with the well-known L\"uscher formalism, using two different implementations of the ERE to describe the finite-volume energy levels. Our findings suggest a non-negligible contribution to the extracted pole properties from the inclusion of the left-hand-cut effects. In all analyses, we found a bound \HD on the physical Riemann sheet very close to the threshold. We also extracted the low-energy parameters of the two-baryon scattering system, \ie, the scattering length and effective range. Finally, the compositeness criterion suggests a potential molecular component within this state. 

\section*{Acknowledgements}
This work uses finite-volume energy levels presented in Ref.~\cite{Green:2021qol}. We thank the authors for useful exchanges on their results. We thank Sebastian~M.~Dawid and Andrew~W.~Jackura for discussions on the $N/D$ formalism and its implementation. We also thank Robert~J.~Perry for discussions on the results and manuscript. This work was supported by the U.S. Department of Energy contract \mbox{DE-AC05-06OR23177}, under which Jefferson Science Associates, LLC operates Jefferson Lab, by U.S.~Department of Energy Grant Nos.~\mbox{DE-FG02-87ER40365}, and \mbox{DE-SC0011090}, and it contributes to the aims of the U.S.~Department of Energy \mbox{ExoHad} Topical Collaboration, contract \mbox{DE-SC0023598}. GM and VM have been supported by the projects \mbox{CEX2024-001451-M} (Unidad de Excelencia ``María de Maeztu''), \mbox{PID2020-118758GB-I00}, and GM additionally by \mbox{PID2023-147112NB-C21}, all financed by \mbox{MICIU/AEI/10.13039/501100011033/} and FEDER, UE, as well as by the EU \mbox{STRONG-2020} project, under the program \mbox{H2020-INFRAIA-2018-1} Grant Agreement \mbox{No.~824093}. VM is a Professor Serra H\'unter. VM acknowledges support from \mbox{CNS2022-136085}. GM was additionally supported by the Beatriu de Pinós program by AGAUR, Grant \mbox{No.~BP 2024 00189}. AP acknowledges support under the program ``Progetti di Rilevante Interesse Nazionale'' (PRIN~2022), published on 2.2.2022 by the Italian Ministry of University and Research (MUR), Project Title ``The X(3872) files'' -- CUP~J53C24002600006 -- Grant Assignment Decree No.~20429 adopted on 6.11.2024 by the Italian Ministry of University and Research (MUR).

\bibliographystyle{elsarticle-num} 
\bibliography{biblio}

\def\includedfrommain{1}
\ifdefined\includedfrommain\else
\documentclass[final,5p,times,twocolumn,nopreprintline]{elsarticle}

\usepackage[T1]{fontenc}
\usepackage[export]{adjustbox}
\usepackage[english]{babel}
\usepackage[utf8]{inputenc}
\usepackage{color,graphicx}
\usepackage{bm,bbm}
\usepackage{enumerate}
\usepackage{hyperref}
\usepackage{mathtools}
\usepackage{mathrsfs}
\usepackage{mciteplus}
\usepackage{slashed}
\usepackage{soul}
\usepackage[normalem]{ulem}
\usepackage{wrapfig}
\usepackage{xspace}
\usepackage[shortlabels]{enumitem}
\usepackage{orcidlink}
\usepackage[capitalise]{cleveref}
\usepackage{txfonts}
\usepackage{relsize}
\usepackage{multirow}
\usepackage{gensymb}
\usepackage{xr-hyper}

\externaldocument{Main}

\hypersetup{
    pdfnewwindow=true,
    colorlinks=true,
    linkcolor=royalblue,
    citecolor=royalblue,
    filecolor=royalblue,
    urlcolor=royalblue
}

\begin{document}
\fi

\clearpage
\onecolumn
\setcounter{section}{0}
\setcounter{figure}{0}
\setcounter{table}{0}
\setcounter{equation}{0}
\renewcommand{\thesection}{S\arabic{section}}
\renewcommand{\thefigure}{S\arabic{figure}}
\renewcommand{\thetable}{S\arabic{table}}
\renewcommand{\theequation}{S\arabic{equation}}
\providecommand{\theHsection}{}
\providecommand{\theHfigure}{}
\providecommand{\theHtable}{}
\providecommand{\theHequation}{}
\renewcommand{\theHsection}{supp.\arabic{section}}
\renewcommand{\theHfigure}{supp.\arabic{figure}}
\renewcommand{\theHtable}{supp.\arabic{table}}
\renewcommand{\theHequation}{supp.\arabic{equation}}

\begin{center}
  {\huge \sc Supplemental Material}
\end{center}

In this Supplemental Material, we provide additional information on the results, including detailed tables of fitted parameters, fits without $a$ dependence, the behavior of the numerator $N(s)$,
and the explicit comparison between the L\"uscher and $N/D$ quantization conditions without left-hand cuts.

\section{Details on fitted parameterizations}
\label{sec:explicit_fits}
In the following, we present detailed results for the explicit fits obtained using \texttt{iminuit}~\cite{James:1975dr,iminuit} included in the Letter, where we consider the average values of the baryon and pion masses over the full set of ensembles. Our masses, energy $\sqrt{s}$, and inverse spacing $1/a$ are all taken in units of GeV. The parameters are given in the appropriate units associated with this choice.

First, the fit using the effective-range expansion (ERE) is presented below. For the fit with $a$ dependence, we use a total of 6 parameters to fit the energy region from the lowest energy level up to the inelastic three-body threshold. We include data points below the left-hand cut in order to compare, on equal footing, with the $N/D$ model. The resulting fit is
\begin{center}
\begin{tabular}{rll}
$c_{00} = $ & $-14.60(71)$ & \multirow{6}{*}{ $\begin{bmatrix*}[r] 1.00 & -0.95 & -0.61 & 0.45 & -0.17 & 0.11 \\
{} & 1.00 & 0.65 & -0.58 & 0.16 & -0.13 \\
{} & {} & 1.00 & -0.91 & 0.58 & -0.58 \\
{} & {} & {} & 1.00 & -0.51 & 0.64 \\
{} & {} & {} & {} & 1.00 & -0.91 \\
{} & {} & {} & {} & {} & 1.00 \end{bmatrix*}$ } \\
$c_{0a} = $ & $25.3(35)$ & \\
$c_{10} = $ & $2.414(77)$ & \\
$c_{1a} = $ & $-1.34(53)$ & \\
$c_{20} = $ & $4.91(80)$ & \\
$c_{2a} = $ & $-6.9(48)$ & \\
\\[1.3ex]
\multicolumn{2}{r}{${\chi^2/N_{\text{dof}}=\frac{41.91}{61-6}=0.76.}$}\\
\end{tabular}
\end{center}

The fit exhibits very good agreement with the data. While some of the resulting parameters, which describe the $a$ dependence, are not compatible with zero, the fit without $a$ dependence also produces a good description of the data and qualitatively similar results, as shown in~\cref{sec:no_a_fits}.

Next, we present the results using ERE with a Chew-Mandelstam (CM) term~\cite{Chew:1960iv}. In this case, the results in the physical region are essentially those of the previous case, with only slight changes in the higher-order parameters of the expansion.

\begin{center}
\begin{tabular}{rll}
$c_{00} = $ & $-14.60(71)$ & \multirow{6}{*}{ $\begin{bmatrix*}[r] 1.00 & -0.95 & -0.61 & 0.45 & -0.17 & 0.11 \\
{} & 1.00 & 0.65 & -0.58 & 0.16 & -0.13 \\
{} & {} & 1.00 & -0.91 & 0.58 & -0.58 \\
{} & {} & {} & 1.00 & -0.51 & 0.64 \\
{} & {} & {} & {} & 1.00 & -0.91 \\
{} & {} & {} & {} & {} & 1.00 \end{bmatrix*}$ } \\
$c_{0a} = $ & $25.3(35)$ & \\
$c_{10} = $ & $1.874(77)$ & \\
$c_{1a} = $ & $-1.34(53)$ & \\
$c_{20} = $ & $4.98(80)$ & \\
$c_{2a} = $ & $-6.8(48)$ & \\
\\[1.3ex]
\multicolumn{2}{r}{${\chi^2/N_{\text{dof}}=\frac{41.92}{61-6}=0.76.}$}\\
\end{tabular}
\end{center}

Once again, fitting with this form when neglecting $a$ dependence produces qualitatively similar results.

Finally, we present below our fit for the $N/D$ model. In this case, we use a total of 8 parameters and obtain a slightly better $\chi^2$ than in the previous cases. Given that the left-hand cut induced by one-pion exchange (OPE) produces small corrections to the predicted energy levels, an extra pair of parameters is needed in order to obtain a $\chi^2$ comparable to those of the ERE models. We note that the spacing-related parameters are closer to zero, within uncertainties, than for the previous models.

\begin{center}
\begin{tabular}{rll}
$n_{00} = $ & $0.280(25)$ & \multirow{8}{*}{ $\begin{bmatrix*}[r] 1.00 & -0.85 & -0.03 & -0.03 & 0.35 & -0.49 & 0.73 & -0.61 \\
{} & 1.00 & -0.08 & 0.09 & -0.52 & 0.34 & -0.39 & 0.84 \\
{} & {} & 1.00 & -0.77 & -0.35 & 0.17 & -0.12 & 0.05 \\
{} & {} & {} & 1.00 & 0.10 & -0.34 & 0.15 & 0.03 \\
{} & {} & {} & {} & 1.00 & -0.37 & -0.30 & -0.38 \\
{} & {} & {} & {} & {} & 1.00 & -0.26 & -0.16 \\
{} & {} & {} & {} & {} & {} & 1.00 & -0.32 \\
{} & {} & {} & {} & {} & {} & {} & 1.00 \end{bmatrix*}$ } \\
$n_{0a} = $ & $0.24(18)$ & \\
$g_{00} = $ & $0.089(25)$ & \\
$g_{0a} = $ & $0.07(14)$ & \\
$d_{00} = $ & $-0.4173(17)$ & \\
$d_{0a} = $ & $0.0342(93)$ & \\
$d_{10} = $ & $0.04780(43)$ & \\
$d_{1a} = $ & $-0.0041(28)$ & \\
\\[1.3ex]
\multicolumn{2}{r}{${\chi^2/N_{\text{dof}}=\frac{40.35}{61-8}=0.76.}$}\\
\end{tabular}
\end{center}

All of these fits produce compatible results in the physical region, as can be seen in~\cref{fig:amp_comparison}. However, they differ significantly below threshold, as expected.

As explained in the main text, one crucial point of the discussion about the $N/D$ method is the presence or absence of Castillejo-Dalitz-Dyson (CDD) poles~\cite{Castillejo:1955ed}, \ie, poles of the denominator $D(s)$, which are not explicitly included in our parameterization. However, they can still be identified by zeros of the numerator above threshold. Without loss of generality, we assume that there is only one simple root $s_z$ above threshold found in the numerator after fitting, i.e., $N(s_z)=0$. As the amplitude does not change if one divides both $N(s)$ and $D(s)$ by a common factor, we can redefine the numerator and denominator as
\begin{align}
\tilde{N}(s)&=\frac{N(s)}{s-s_z},\\
\tilde{D}(s)&=\frac{D(s)}{s-s_z}=\frac{\tilde{d}_{-1}}{s-s_z}+\sum_{j=0}^{n-1}\tilde{d}_j s^j+\frac{s}{\pi(s-s_z)}\int_{s_{\text{thr}}}^{\infty}ds'\frac{\rho(s')N(s')}{s'(s-s')},
\end{align}
where $\tilde{d}_j$ are polynomials in $s_z$ and the expansion coefficients $d_j$ of $D(s)$, and $\tilde{N}(s)$ is now free of zeros. Using the relation,
\begin{align}
\frac{N(s')}{s-s_z}&=\frac{N(s')}{s'-s_z}+\frac{N(s')(s'-s)}{(s'-s_z)(s-s_z)}=\tilde{N}(s')+\frac{s'-s}{s-s_z}\tilde{N}(s'),
\end{align}
we have
\begin{align}
\tilde{D}(s)&=\frac{\tilde{d}_{-1}}{s-s_z}+\tilde{d}_0+\sum_{j=1}^{n-1}\tilde{d}_j s^j+\frac{s}{\pi}\int_{s_{\text{thr}}}^{\infty}ds' \frac{\rho(s')\tilde{N}(s')}{s'(s-s')}-\frac{s}{\pi(s-s_z)}\int_{s_{\text{thr}}}^{\infty}ds' \frac{\rho(s')\tilde{N}(s')}{s'},
\end{align}
where the second integral is convergent and can be absorbed into the first two terms. Finally, we obtain
\begin{align}
\tilde{D}(s)=\frac{\tilde{d}'_{-1}}{s-s_z}+\tilde{d}'_0+\sum_{j=1}^{n-1}\tilde{d}_j s^j+\frac{s}{\pi}\int_{s_{\text{thr}}}^{\infty}ds' \frac{\rho(s')\tilde{N}(s')}{s'(s-s')},
\end{align}
where we can further normalize $\tilde{D}(s)$ by dividing both the new numerator and denominator by $\tilde{d}'_0$, and thus we conclude that the need for additional CDD poles can be accommodated by the zeros of the numerator.

\begin{figure*}
\centering
\includegraphics[width=0.9\textwidth]{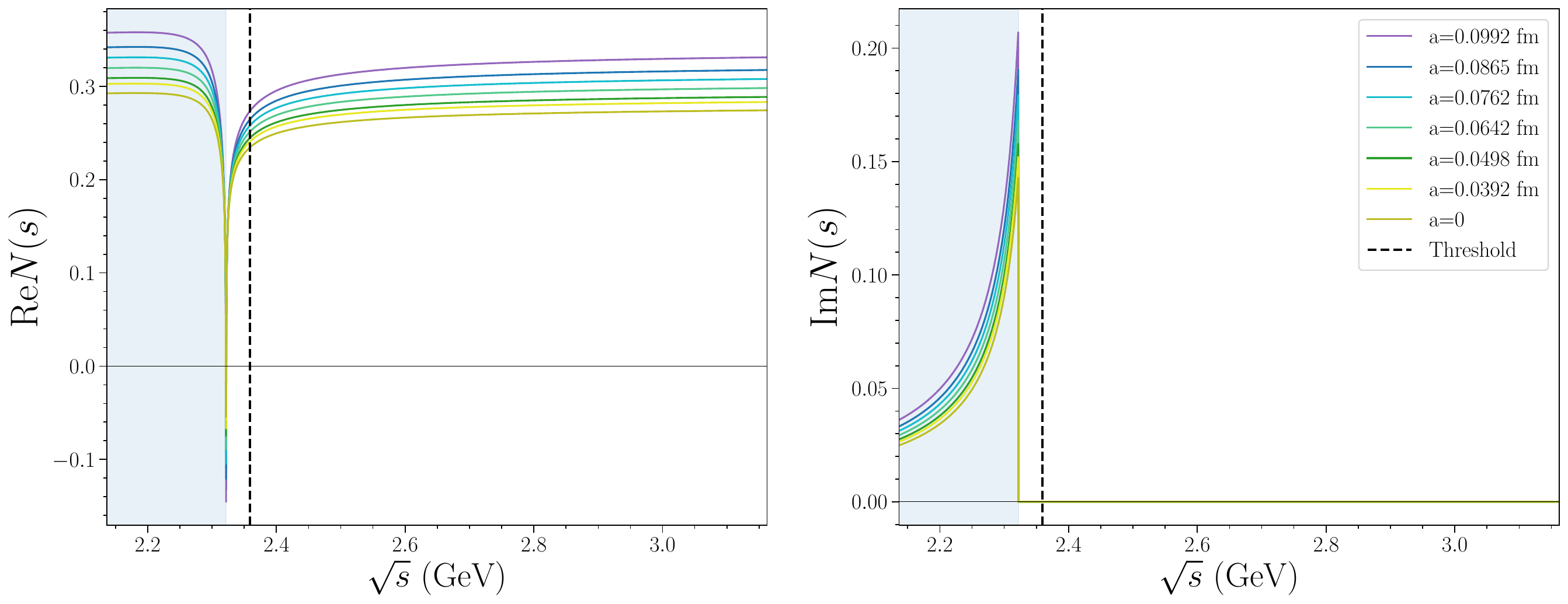}
\caption{Real (left) and imaginary (right) parts of the numerator at each spacing $a$ of the lattices used in this Letter. Note that these numerators do not cross zero above threshold. 
}
\label{fig:N_comparison}
\end{figure*}

In~\cref{fig:N_comparison}, we show the different numerators obtained from our fits to the data for each ensemble, identified by its corresponding spacing. There is no zero crossing above the vicinity of the left-hand cut, which hints at the absence of any CDD pole. In fact, the resulting numerator in the continuum is described by only two parameters and is smooth in the physical region. It is not dissimilar to a constant value above threshold, which would correspond to the ERE model with a Chew-Mandelstam term. This explains why the phase shifts in the physical region are almost identical among the different models, while the behavior below threshold is different. The OPE effect distinguishes the $N/D$ model from the ERE parameterizations there.

\section{Fits without $a$ dependence}
\label{sec:no_a_fits}

\begin{figure*}
\centering
\includegraphics[width=0.9\textwidth]{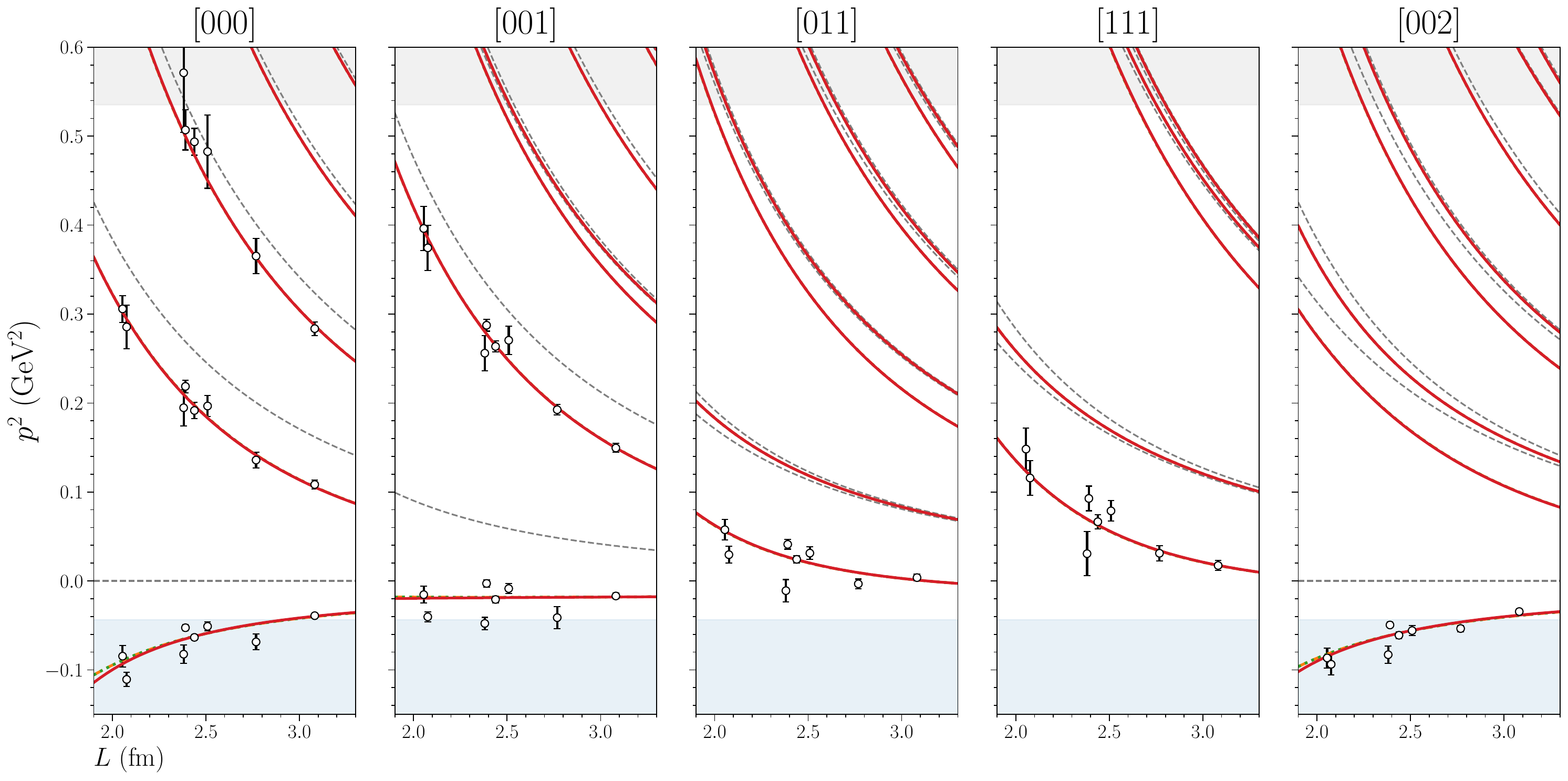}
\caption{Similar to~\cref{fig:spectrum_a},
we show a comparison of the spectra obtained by our three different models, fitted to the finite-volume data while neglecting any $a$ dependence. The red line shows the $N/D$ model fit, while the orange dashed line presents the ERE fit, and the green dotted line shows the ERE with a Chew-Mandelstam term fit. No fitting uncertainties are shown for clarity of presentation. Grey dashed lines display the non-interacting energy levels. The grey shaded area represents the inelastic region where three-body effects may become relevant~\cite{Briceno:2018aml}, and the blue area represents the region where the OPE left-hand cut is located.}
\label{fig:spectrum}
\end{figure*}

In this section, we neglect any $a$ dependence when describing the two-body energy levels, fix the baryon mass to the average over the various lattices in Ref.~\cite{Green:2021qol}, and repeat all the analyses presented in the Letter. Our goal is twofold. First, the prescription used in this work is not a rigorous approach to include $a$-dependent effects in the finite-volume formalisms, but rather a phenomenological parameterization. We would like to check whether there is a significant difference between including and neglecting these effects. Second, we would like to test whether the differences between the $N/D$ and ERE parameterizations still manifest if we ignore the $a$ dependence.

We find that fits neglecting $a$ dependence still produce a qualitative description of the data. As in~\cref{sec:explicit_fits}, we repeat the analysis and show only statistical uncertainties here. The ERE parameterizations, without and with the Chew-Mandelstam term (left and right tables below, respectively), produce once again almost identical fit results.

\begin{table*}[!h]
\centering
\begin{minipage}[t]{0.48\textwidth}
\begin{center}
\begin{tabular}{rll}
$c_{0} = $ & $-10.94(26)$ & \multirow{3}{*}{ $\begin{bmatrix*}[r] 1.00 & -0.68 & -0.30 \\
{} & 1.00 & 0.67 \\
{} & {} & 1.00 \end{bmatrix*}$ } \\
$c_{1} = $ & $2.219(31)$ & \\
$c_{2} = $ & $3.85(33)$ & \\
\\[1.3ex]
\multicolumn{2}{r}{${\chi^2/N_{\text{dof}}=\frac{95.83}{61-3}=1.65.}$}\\
\end{tabular}
\end{center}

\end{minipage}
\hfill
\begin{minipage}[t]{0.48\textwidth}
\begin{center}
\begin{tabular}{rll}
$c_{0} = $ & $-10.94(26)$ & \multirow{3}{*}{ $\begin{bmatrix*}[r] 1.00 & -0.68 & -0.30 \\
{} & 1.00 & 0.67 \\
{} & {} & 1.00 \end{bmatrix*}$ } \\
$c_{1} = $ & $1.679(31)$ & \\
$c_{2} = $ & $3.91(33)$ & \\
\\[1.3ex]
\multicolumn{2}{r}{${\chi^2/N_{\text{dof}}=\frac{95.83}{61-3}=1.65.}$}\\
\end{tabular}
\end{center}

\end{minipage}
\end{table*}

Similarly, if an extra parameter is included, the $N/D$ model produces slightly better results once again. Furthermore, these parameters are closer to the spacing-dependent parameters above than are the ERE parameters.

\begin{center}
\begin{tabular}{rll}
$n_0 = $ & $0.317(12)$ & \multirow{4}{*}{ $\begin{bmatrix*}[r] 1.00 & -0.40 & -0.18 & 0.83 \\
{} & 1.00 & -0.57 & -0.11 \\
{} & {} & 1.00 & -0.67 \\
{} & {} & {} & 1.00 \end{bmatrix*}$ } \\
$g_0 = $ & $0.097(17)$ & \\
$d_{0} = $ & $-0.4127(13)$ & \\
$d_{1} = $ & $0.04734(35)$ & \\
\\[1.3ex]
\multicolumn{2}{r}{${\chi^2/N_{\text{dof}}=\frac{94.30}{61-4}=1.65.}$}\\
\end{tabular}
\end{center}

\begin{figure}
  \centering
\includegraphics[width=0.9\textwidth]{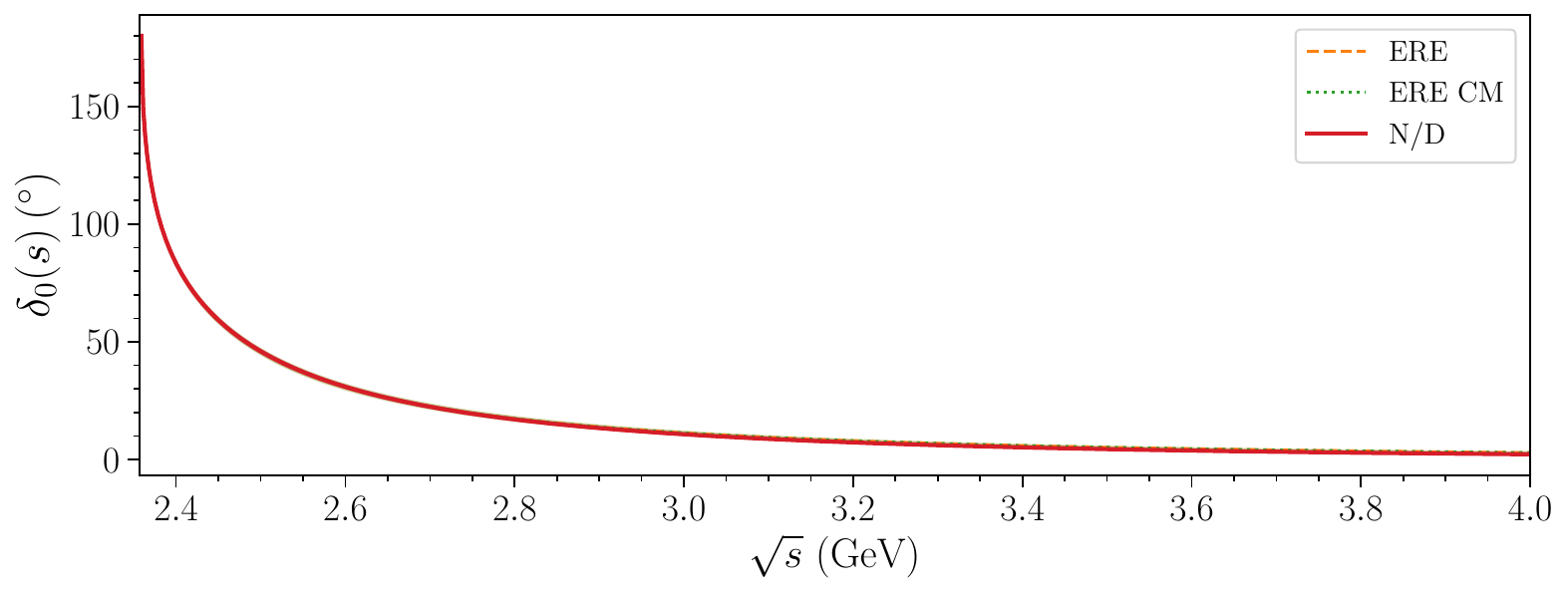}
\caption{Comparison of the three different phase shifts predicted in the physical region by each of the models fitted to the finite-volume data. Note that the fits shown in this plot neglect any spacing dependence.}
\label{fig:ps}
\end{figure}

In~\cref{fig:spectrum}, we compare the data description among the three models, using the same colors as in the main text. In this case, the $N/D$ formalism describes the ground-state levels slightly better than the ERE ones. For all other levels, the three models produce almost identical results.

Note that the ground state exhibits a strong volume dependence, which is not surprising if one considers this level to be dominated by a very shallow bound state. In this case, however, the state appears to be more bound than before, but the finite-volume amplitudes exhibit an even greater dependence on the volume than when including $a$-dependent terms.

In~\cref{fig:ps}, the phase shifts predicted in the physical region are shown. They are similar to those presented in the main text. Their line shapes decrease sharply from 180$\degree$ at threshold and approach 0$\degree$ quickly, which implies that the partial wave amplitudes approach zero rapidly away from threshold. 

\begin{figure*}[!htb]
\centering
\includegraphics[width=0.8\textwidth]{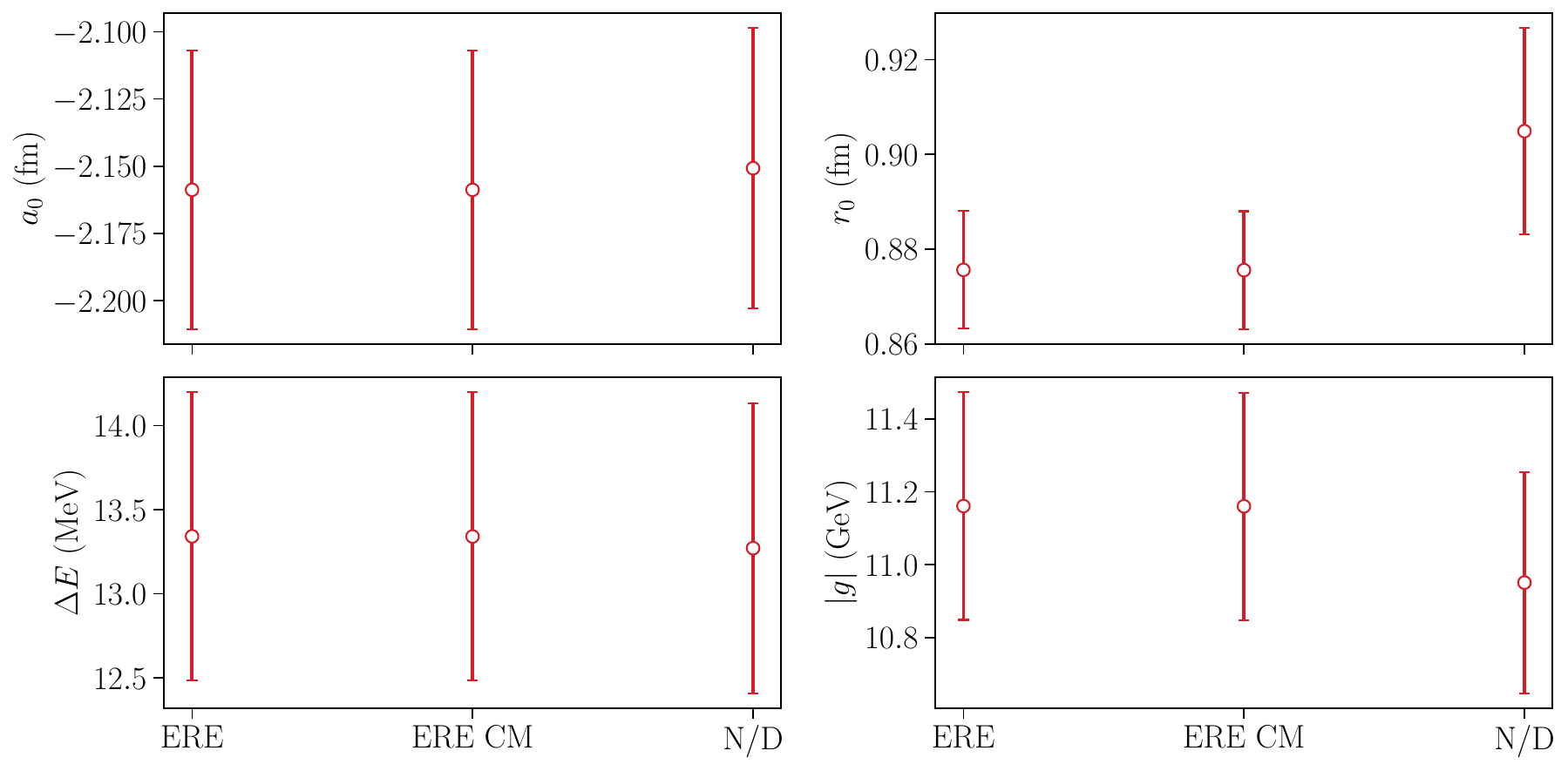}
\caption{Observables extracted from the three different models used to describe the finite-volume data neglecting spacing dependence. Top: scattering length (left) and effective range (right). Bottom: binding energy (left) and modulus of the coupling (right).}
\label{fig:poles}
\end{figure*} 

In the top panel of~\cref{fig:poles}, the scattering lengths and effective ranges extracted from fits without spacing dependence are plotted. Both absolute values decrease due to the absorption of the $a$-dependent effects. The behavior around threshold strongly hints at the existence of a bound state nearby. Following the procedure described in the main text, we can once again extract the pole positions located on the real axis below threshold, presented in the bottom panel of~\cref{fig:poles}.

Compared to the results in~\cref{fig:poles_a}, the poles are more bound and thus incompatible with those results. In particular, the $N/D$ model has the largest shift in value. As remarked before, the $a$-dependent analysis here is not a rigorous approach, as no such framework exists. These deviations could be explained by the fact that $a$-dependent fits include several weakly constrained parameters, affecting some observables more than others.

Nevertheless, it is worth noting that all these results agree with the claim that the \HD appears as a bound state for these lattices, regardless of the left-hand cuts or $a$-dependent effects.

\section{Equivalence between $N/D$ and L\"uscher formalisms}
\label{sec:equivalence}
As detailed in Ref.~\cite{Dawid:2024oey}, the finite-volume $N/D$ formalism is equivalent to the L\"uscher quantization condition for energies above the left-hand cut. In their derivation, the authors made use of the fact that $N$ is identical above threshold in finite and infinite volumes, up to exponentially suppressed corrections. Therefore, by virtue of the sum-integral difference~\cite{Kim:2005gf,Briceno:2015tza}, they identified the quantization condition $D_{\text{FV}}=0$ with the one expressed by the $K$-matrix, which is just the principal value of $D_{\text{IFV}}$ rescaled by the numerator, along with another term corresponding exactly to~\cref{eq:Fpv} in the main text. In so doing, the $N/D$ method can be rewritten as the L\"uscher formalism above the energy region where left-hand cuts are present.

In the following, we illustrate this equivalence within a simplified $N/D$ model where left-hand cuts are turned off, showing consistency between both quantization conditions in the entire energy region under study. The model presented here is still based on a once-subtracted dispersive representation, where
\begin{align}
N(s)=1.
\end{align}

\begin{figure*}[!h]
\centering
\includegraphics[width=0.9\textwidth]{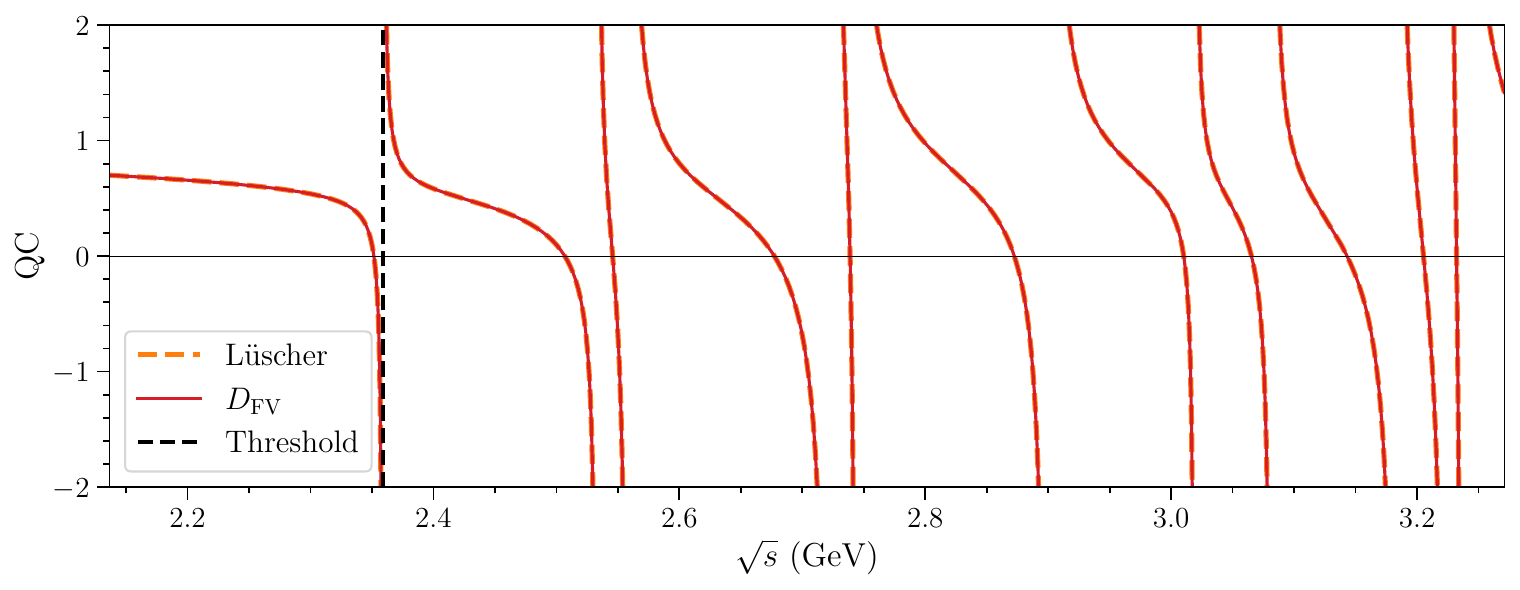}
\caption{Comparison between the L\"uscher and $N/D$ quantization conditions (QCs) for a once-subtracted model with $N(s)=1$. In this comparison, we use $m_B=1.18\gev$ for a lattice with $L=2.5$ fm and $\vec n=[2,0,0]$.}
\label{fig:qc_comparison}
\end{figure*}

As shown in~\cref{fig:qc_comparison}, where we compare both quantization conditions when $N(s)=1$, the results match very well, up to exponentially suppressed corrections. When $N(s)$ is a function different from 1, $D_{\text{FV}}$ is identical to the L\"uscher equation times $N(s)$, which produces the same zeros for both methods.

\ifdefined\includedfrommain\else
\bibliographystyle{elsarticle-num}
\bibliography{biblio}

\begin{thebibliography}{10}
\expandafter\ifx\csname url\endcsname\relax
  \def\url#1{\texttt{#1}}\fi
\expandafter\ifx\csname urlprefix\endcsname\relax\def\urlprefix{URL }\fi
\expandafter\ifx\csname href\endcsname\relax
  \def\href#1#2{#2} \def\path#1{#1}\fi

\bibitem{Belle:2003nnu}
S.~K. Choi, et~al., {Observation of a narrow charmonium-like state in exclusive $B^\pm \to K^\pm \pi^+ \pi^- J/\psi$ decays}, Phys. Rev. Lett. 91 (2003) 262001.
\newblock \href {http://arxiv.org/abs/hep-ex/0309032} {\path{arXiv:hep-ex/0309032}}, \href {https://doi.org/10.1103/PhysRevLett.91.262001} {\path{doi:10.1103/PhysRevLett.91.262001}}.

\bibitem{BaBar:2004oro}
B.~Aubert, et~al., {Study of the $B \to J/\psi K^- \pi^+ \pi^-$ decay and measurement of the $B \to X(3872) K^-$ branching fraction}, Phys. Rev. D 71 (2005) 071103.
\newblock \href {http://arxiv.org/abs/hep-ex/0406022} {\path{arXiv:hep-ex/0406022}}, \href {https://doi.org/10.1103/PhysRevD.71.071103} {\path{doi:10.1103/PhysRevD.71.071103}}.

\bibitem{LHCb:2021vvq}
R.~Aaij, et~al., {Observation of an exotic narrow doubly charmed tetraquark}, Nature Phys. 18~(7) (2022) 751--754.
\newblock \href {http://arxiv.org/abs/2109.01038} {\path{arXiv:2109.01038}}, \href {https://doi.org/10.1038/s41567-022-01614-y} {\path{doi:10.1038/s41567-022-01614-y}}.

\bibitem{LHCb:2021auc}
R.~Aaij, et~al., {Study of the doubly charmed tetraquark $T_{cc}^{+}$}, Nature Commun. 13~(1) (2022) 3351.
\newblock \href {http://arxiv.org/abs/2109.01056} {\path{arXiv:2109.01056}}, \href {https://doi.org/10.1038/s41467-022-30206-w} {\path{doi:10.1038/s41467-022-30206-w}}.

\bibitem{LHCb:2015yax}
R.~Aaij, et~al., {Observation of $J/\psi p$ Resonances Consistent with Pentaquark States in $\Lambda_b^0 \to J/\psi K^- p$ Decays}, Phys. Rev. Lett. 115 (2015) 072001.
\newblock \href {http://arxiv.org/abs/1507.03414} {\path{arXiv:1507.03414}}, \href {https://doi.org/10.1103/PhysRevLett.115.072001} {\path{doi:10.1103/PhysRevLett.115.072001}}.

\bibitem{LHCb:2019kea}
R.~Aaij, et~al., {Observation of a narrow pentaquark state, $P_c(4312)^+$, and of two-peak structure of the $P_c(4450)^+$}, Phys. Rev. Lett. 122~(22) (2019) 222001.
\newblock \href {http://arxiv.org/abs/1904.03947} {\path{arXiv:1904.03947}}, \href {https://doi.org/10.1103/PhysRevLett.122.222001} {\path{doi:10.1103/PhysRevLett.122.222001}}.

\bibitem{JPAC:2018zyd}
A.~Rodas, et~al., {Determination of the pole position of the lightest hybrid meson candidate}, Phys. Rev. Lett. 122~(4) (2019) 042002.
\newblock \href {http://arxiv.org/abs/1810.04171} {\path{arXiv:1810.04171}}, \href {https://doi.org/10.1103/PhysRevLett.122.042002} {\path{doi:10.1103/PhysRevLett.122.042002}}.

\bibitem{GlueX:2024erj}
F.~Afzal, et~al., {Upper Limit on the Photoproduction Cross Section of the Spin-Exotic {\ensuremath{\pi}}1(1600)}, Phys. Rev. Lett. 133~(26) (2024) 261903.
\newblock \href {http://arxiv.org/abs/2407.03316} {\path{arXiv:2407.03316}}, \href {https://doi.org/10.1103/PhysRevLett.133.261903} {\path{doi:10.1103/PhysRevLett.133.261903}}.

\bibitem{BESIII:2022riz}
M.~Ablikim, et~al., {Observation of an Isoscalar Resonance with Exotic JPC=1-+ Quantum Numbers in J/{\ensuremath{\psi}}{\textrightarrow}{\ensuremath{\gamma}}{\ensuremath{\eta}}{\ensuremath{\eta}}'}, Phys. Rev. Lett. 129~(19) (2022) 192002, [Erratum: Phys.Rev.Lett. 130, 159901 (2023)].
\newblock \href {http://arxiv.org/abs/2202.00621} {\path{arXiv:2202.00621}}, \href {https://doi.org/10.1103/PhysRevLett.129.192002} {\path{doi:10.1103/PhysRevLett.129.192002}}.

\bibitem{Guo:2017jvc}
F.-K. Guo, C.~Hanhart, U.-G. Mei{\ss}ner, Q.~Wang, Q.~Zhao, B.-S. Zou, {Hadronic molecules}, Rev. Mod. Phys. 90~(1) (2018) 015004, [Erratum: Rev.Mod.Phys. 94, 029901 (2022)].
\newblock \href {http://arxiv.org/abs/1705.00141} {\path{arXiv:1705.00141}}, \href {https://doi.org/10.1103/RevModPhys.90.015004} {\path{doi:10.1103/RevModPhys.90.015004}}.

\bibitem{Chen:2016qju}
H.-X. Chen, W.~Chen, X.~Liu, S.-L. Zhu, {The hidden-charm pentaquark and tetraquark states}, Phys. Rept. 639 (2016) 1--121.
\newblock \href {http://arxiv.org/abs/1601.02092} {\path{arXiv:1601.02092}}, \href {https://doi.org/10.1016/j.physrep.2016.05.004} {\path{doi:10.1016/j.physrep.2016.05.004}}.

\bibitem{Chen:2022asf}
H.-X. Chen, W.~Chen, X.~Liu, Y.-R. Liu, S.-L. Zhu, {An updated review of the new hadron states}, Rept. Prog. Phys. 86~(2) (2023) 026201.
\newblock \href {http://arxiv.org/abs/2204.02649} {\path{arXiv:2204.02649}}, \href {https://doi.org/10.1088/1361-6633/aca3b6} {\path{doi:10.1088/1361-6633/aca3b6}}.

\bibitem{Brambilla:2019esw}
N.~Brambilla, S.~Eidelman, C.~Hanhart, A.~Nefediev, C.-P. Shen, C.~E. Thomas, A.~Vairo, C.-Z. Yuan, {The $XYZ$ states: experimental and theoretical status and perspectives}, Phys. Rept. 873 (2020) 1--154.
\newblock \href {http://arxiv.org/abs/1907.07583} {\path{arXiv:1907.07583}}, \href {https://doi.org/10.1016/j.physrep.2020.05.001} {\path{doi:10.1016/j.physrep.2020.05.001}}.

\bibitem{Pelaez:2025wma}
J.~R. Pelaez, {Light Meson Resonances} (9 2025).
\newblock \href {http://arxiv.org/abs/2509.08648} {\path{arXiv:2509.08648}}.

\bibitem{Tornqvist:1991ks}
N.~A. Tornqvist, {Possible large deuteron - like meson meson states bound by pions}, Phys. Rev. Lett. 67 (1991) 556--559.
\newblock \href {https://doi.org/10.1103/PhysRevLett.67.556} {\path{doi:10.1103/PhysRevLett.67.556}}.

\bibitem{Tornqvist:1993ng}
N.~A. Tornqvist, {From the deuteron to deusons, an analysis of deuteron - like meson meson bound states}, Z. Phys. C 61 (1994) 525--537.
\newblock \href {http://arxiv.org/abs/hep-ph/9310247} {\path{arXiv:hep-ph/9310247}}, \href {https://doi.org/10.1007/BF01413192} {\path{doi:10.1007/BF01413192}}.

\bibitem{Wang:2013kva}
P.~Wang, X.~G. Wang, {Study on X(3872) from effective field theory with pion exchange interaction}, Phys. Rev. Lett. 111~(4) (2013) 042002.
\newblock \href {http://arxiv.org/abs/1304.0846} {\path{arXiv:1304.0846}}, \href {https://doi.org/10.1103/PhysRevLett.111.042002} {\path{doi:10.1103/PhysRevLett.111.042002}}.

\bibitem{Karliner:2015ina}
M.~Karliner, J.~L. Rosner, {New Exotic Meson and Baryon Resonances from Doubly-Heavy Hadronic Molecules}, Phys. Rev. Lett. 115~(12) (2015) 122001.
\newblock \href {http://arxiv.org/abs/1506.06386} {\path{arXiv:1506.06386}}, \href {https://doi.org/10.1103/PhysRevLett.115.122001} {\path{doi:10.1103/PhysRevLett.115.122001}}.

\bibitem{Baru:2015nea}
V.~Baru, E.~Epelbaum, A.~A. Filin, F.~K. Guo, H.~W. Hammer, C.~Hanhart, U.~G. Mei{\ss}ner, A.~V. Nefediev, {Remarks on study of X(3872) from effective field theory with pion-exchange interaction}, Phys. Rev. D 91~(3) (2015) 034002.
\newblock \href {http://arxiv.org/abs/1501.02924} {\path{arXiv:1501.02924}}, \href {https://doi.org/10.1103/PhysRevD.91.034002} {\path{doi:10.1103/PhysRevD.91.034002}}.

\bibitem{PavonValderrama:2019nbk}
M.~Pavon~Valderrama, {One pion exchange and the quantum numbers of the P$_c$(4440) and P$_c$(4457) pentaquarks}, Phys. Rev. D 100~(9) (2019) 094028.
\newblock \href {http://arxiv.org/abs/1907.05294} {\path{arXiv:1907.05294}}, \href {https://doi.org/10.1103/PhysRevD.100.094028} {\path{doi:10.1103/PhysRevD.100.094028}}.

\bibitem{Du:2023hlu}
M.-L. Du, A.~Filin, V.~Baru, X.-K. Dong, E.~Epelbaum, F.-K. Guo, C.~Hanhart, A.~Nefediev, J.~Nieves, Q.~Wang, {Role of Left-Hand Cut Contributions on Pole Extractions from Lattice Data: Case Study for Tcc(3875)+}, Phys. Rev. Lett. 131~(13) (2023) 131903.
\newblock \href {http://arxiv.org/abs/2303.09441} {\path{arXiv:2303.09441}}, \href {https://doi.org/10.1103/PhysRevLett.131.131903} {\path{doi:10.1103/PhysRevLett.131.131903}}.

\bibitem{Luscher:1986pf}
M.~Luscher, {Volume Dependence of the Energy Spectrum in Massive Quantum Field Theories. 2. Scattering States}, Commun. Math. Phys. 105 (1986) 153--188.
\newblock \href {https://doi.org/10.1007/BF01211097} {\path{doi:10.1007/BF01211097}}.

\bibitem{Luscher:1990ux}
M.~Luscher, {Two particle states on a torus and their relation to the scattering matrix}, Nucl. Phys. B 354 (1991) 531--578.
\newblock \href {https://doi.org/10.1016/0550-3213(91)90366-6} {\path{doi:10.1016/0550-3213(91)90366-6}}.

\bibitem{Briceno:2017max}
R.~A. Briceno, J.~J. Dudek, R.~D. Young, {Scattering processes and resonances from lattice QCD}, Rev. Mod. Phys. 90~(2) (2018) 025001.
\newblock \href {http://arxiv.org/abs/1706.06223} {\path{arXiv:1706.06223}}, \href {https://doi.org/10.1103/RevModPhys.90.025001} {\path{doi:10.1103/RevModPhys.90.025001}}.

\bibitem{Rummukainen:1995vs}
K.~Rummukainen, S.~A. Gottlieb, {Resonance scattering phase shifts on a nonrest frame lattice}, Nucl. Phys. B 450 (1995) 397--436.
\newblock \href {http://arxiv.org/abs/hep-lat/9503028} {\path{arXiv:hep-lat/9503028}}, \href {https://doi.org/10.1016/0550-3213(95)00313-H} {\path{doi:10.1016/0550-3213(95)00313-H}}.

\bibitem{Kim:2005gf}
C.~h. Kim, C.~T. Sachrajda, S.~R. Sharpe, {Finite-volume effects for two-hadron states in moving frames}, Nucl. Phys. B 727 (2005) 218--243.
\newblock \href {http://arxiv.org/abs/hep-lat/0507006} {\path{arXiv:hep-lat/0507006}}, \href {https://doi.org/10.1016/j.nuclphysb.2005.08.029} {\path{doi:10.1016/j.nuclphysb.2005.08.029}}.

\bibitem{Leskovec:2012gb}
L.~Leskovec, S.~Prelovsek, {Scattering phase shifts for two particles of different mass and non-zero total momentum in lattice QCD}, Phys. Rev. D 85 (2012) 114507.
\newblock \href {http://arxiv.org/abs/1202.2145} {\path{arXiv:1202.2145}}, \href {https://doi.org/10.1103/PhysRevD.85.114507} {\path{doi:10.1103/PhysRevD.85.114507}}.

\bibitem{Briceno:2012yi}
R.~A. Briceno, Z.~Davoudi, {Moving multichannel systems in a finite volume with application to proton-proton fusion}, Phys. Rev. D 88~(9) (2013) 094507.
\newblock \href {http://arxiv.org/abs/1204.1110} {\path{arXiv:1204.1110}}, \href {https://doi.org/10.1103/PhysRevD.88.094507} {\path{doi:10.1103/PhysRevD.88.094507}}.

\bibitem{Briceno:2014oea}
R.~A. Briceno, {Two-particle multichannel systems in a finite volume with arbitrary spin}, Phys. Rev. D 89~(7) (2014) 074507.
\newblock \href {http://arxiv.org/abs/1401.3312} {\path{arXiv:1401.3312}}, \href {https://doi.org/10.1103/PhysRevD.89.074507} {\path{doi:10.1103/PhysRevD.89.074507}}.

\bibitem{Meng:2021uhz}
L.~Meng, E.~Epelbaum, {Two-particle scattering from finite-volume quantization conditions using the plane wave basis}, JHEP 10 (2021) 051.
\newblock \href {http://arxiv.org/abs/2108.02709} {\path{arXiv:2108.02709}}, \href {https://doi.org/10.1007/JHEP10(2021)051} {\path{doi:10.1007/JHEP10(2021)051}}.

\bibitem{Meng:2023bmz}
L.~Meng, V.~Baru, E.~Epelbaum, A.~A. Filin, A.~M. Gasparyan, {Solving the left-hand cut problem in lattice QCD: Tcc(3875)+ from finite volume energy levels}, Phys. Rev. D 109~(7) (2024) L071506.
\newblock \href {http://arxiv.org/abs/2312.01930} {\path{arXiv:2312.01930}}, \href {https://doi.org/10.1103/PhysRevD.109.L071506} {\path{doi:10.1103/PhysRevD.109.L071506}}.

\bibitem{Raposo:2023oru}
A.~B. Raposo, M.~T. Hansen, {Finite-volume scattering on the left-hand cut}, JHEP 08 (2024) 075.
\newblock \href {http://arxiv.org/abs/2311.18793} {\path{arXiv:2311.18793}}, \href {https://doi.org/10.1007/JHEP08(2024)075} {\path{doi:10.1007/JHEP08(2024)075}}.

\bibitem{Bubna:2024izx}
R.~Bubna, H.-W. Hammer, F.~M{\"u}ller, J.-Y. Pang, A.~Rusetsky, J.-J. Wu, {L{\"u}scher equation with long-range forces}, JHEP 05 (2024) 168.
\newblock \href {http://arxiv.org/abs/2402.12985} {\path{arXiv:2402.12985}}, \href {https://doi.org/10.1007/JHEP05(2024)168} {\path{doi:10.1007/JHEP05(2024)168}}.

\bibitem{Bubna:2025gsd}
R.~Bubna, H.-W. Hammer, B.-L. Hoid, J.-Y. Pang, A.~Rusetsky, J.-J. Wu, {Modified L{\"u}scher zeta-function and the modified effective range expansion in the presence of a long-range force}, JHEP 10 (2025) 197.
\newblock \href {http://arxiv.org/abs/2507.18399} {\path{arXiv:2507.18399}}, \href {https://doi.org/10.1007/JHEP10(2025)197} {\path{doi:10.1007/JHEP10(2025)197}}.

\bibitem{Raposo:2025dkb}
A.~B. Raposo, R.~A. Brice{\~n}o, M.~T. Hansen, A.~W. Jackura, {Extracting scattering amplitudes for arbitrary two-particle systems with one-particle left-hand cuts via lattice QCD}, JHEP 06 (2025) 186.
\newblock \href {http://arxiv.org/abs/2502.19375} {\path{arXiv:2502.19375}}, \href {https://doi.org/10.1007/JHEP06(2025)186} {\path{doi:10.1007/JHEP06(2025)186}}.

\bibitem{Hansen:2024ffk}
M.~T. Hansen, F.~Romero-L{\'o}pez, S.~R. Sharpe, {Incorporating DD{\ensuremath{\pi}} effects and left-hand cuts in lattice QCD studies of the T$_{cc}$(3875)$^{+}$}, JHEP 06 (2024) 051.
\newblock \href {http://arxiv.org/abs/2401.06609} {\path{arXiv:2401.06609}}, \href {https://doi.org/10.1007/JHEP06(2024)051} {\path{doi:10.1007/JHEP06(2024)051}}.

\bibitem{Dawid:2024dgy}
S.~M. Dawid, F.~Romero-L{\'o}pez, S.~R. Sharpe, {Finite- and infinite-volume study of DD{\ensuremath{\pi}} scattering}, JHEP 01 (2025) 060.
\newblock \href {http://arxiv.org/abs/2409.17059} {\path{arXiv:2409.17059}}, \href {https://doi.org/10.1007/JHEP01(2025)060} {\path{doi:10.1007/JHEP01(2025)060}}.

\bibitem{Dawid:2024oey}
S.~M. Dawid, A.~W. Jackura, A.~P. Szczepaniak, {Finite-volume quantization condition from the N/D representation}, Phys. Lett. B 864 (2025) 139442.
\newblock \href {http://arxiv.org/abs/2411.15730} {\path{arXiv:2411.15730}}, \href {https://doi.org/10.1016/j.physletb.2025.139442} {\path{doi:10.1016/j.physletb.2025.139442}}.

\bibitem{Chew:1960iv}
G.~F. Chew, S.~Mandelstam, {Theory of low-energy pion pion interactions}, Phys. Rev. 119 (1960) 467--477.
\newblock \href {https://doi.org/10.1103/PhysRev.119.467} {\path{doi:10.1103/PhysRev.119.467}}.

\bibitem{Bjorken:1960zz}
J.~D. Bjorken, {Construction of Coupled Scattering and Production Amplitudes Satisfying Analyticity and Unitarity}, Phys. Rev. Lett. 4 (1960) 473--474.
\newblock \href {https://doi.org/10.1103/PhysRevLett.4.473} {\path{doi:10.1103/PhysRevLett.4.473}}.

\bibitem{Jaffe:1976yi}
R.~L. Jaffe, {Perhaps a Stable Dihyperon}, Phys. Rev. Lett. 38 (1977) 195--198, [Erratum: Phys.Rev.Lett. 38, 617 (1977)].
\newblock \href {https://doi.org/10.1103/PhysRevLett.38.195} {\path{doi:10.1103/PhysRevLett.38.195}}.

\bibitem{Gross:2018ivp}
C.~Gross, A.~Polosa, A.~Strumia, A.~Urbano, W.~Xue, {Dark Matter in the Standard Model?}, Phys. Rev. D 98~(6) (2018) 063005.
\newblock \href {http://arxiv.org/abs/1803.10242} {\path{arXiv:1803.10242}}, \href {https://doi.org/10.1103/PhysRevD.98.063005} {\path{doi:10.1103/PhysRevD.98.063005}}.

\bibitem{Farrar:2020zeo}
G.~R. Farrar, Z.~Wang, X.~Xu, {Dark Matter Particle in QCD} (7 2020).
\newblock \href {http://arxiv.org/abs/2007.10378} {\path{arXiv:2007.10378}}.

\bibitem{Takahashi:2001nm}
H.~Takahashi, et~al., {Observation of a (Lambda Lambda)He-6 double hypernucleus}, Phys. Rev. Lett. 87 (2001) 212502.
\newblock \href {https://doi.org/10.1103/PhysRevLett.87.212502} {\path{doi:10.1103/PhysRevLett.87.212502}}.

\bibitem{KEKE176:2009jzw}
S.~Aoki, et~al., {Nuclear capture at rest of $\Xi^-$ hyperons}, Nucl. Phys. A 828 (2009) 191--232.
\newblock \href {https://doi.org/10.1016/j.nuclphysa.2009.07.005} {\path{doi:10.1016/j.nuclphysa.2009.07.005}}.

\bibitem{E373KEK-PS:2013dfg}
J.~K. Ahn, et~al., {Double-$\Lambda$ hypernuclei observed in a hybrid emulsion experiment}, Phys. Rev. C 88~(1) (2013) 014003.
\newblock \href {https://doi.org/10.1103/PhysRevC.88.014003} {\path{doi:10.1103/PhysRevC.88.014003}}.

\bibitem{Belle:2013sba}
B.~H. Kim, et~al., {Search for an $H$-dibaryon with mass near $2m_\Lambda$ in $\Upsilon(1S)$ and $\Upsilon(2S)$ decays}, Phys. Rev. Lett. 110~(22) (2013) 222002.
\newblock \href {http://arxiv.org/abs/1302.4028} {\path{arXiv:1302.4028}}, \href {https://doi.org/10.1103/PhysRevLett.110.222002} {\path{doi:10.1103/PhysRevLett.110.222002}}.

\bibitem{STAR:2014dcy}
L.~Adamczyk, et~al., {$\Lambda\Lambda$ Correlation Function in Au+Au collisions at $\sqrt{s_{NN}}=$ 200 GeV}, Phys. Rev. Lett. 114~(2) (2015) 022301.
\newblock \href {http://arxiv.org/abs/1408.4360} {\path{arXiv:1408.4360}}, \href {https://doi.org/10.1103/PhysRevLett.114.022301} {\path{doi:10.1103/PhysRevLett.114.022301}}.

\bibitem{ALICE:2019eol}
S.~Acharya, et~al., {Study of the $\Lambda$-$\Lambda$ interaction with femtoscopy correlations in pp and p-Pb collisions at the LHC}, Phys. Lett. B 797 (2019) 134822.
\newblock \href {http://arxiv.org/abs/1905.07209} {\path{arXiv:1905.07209}}, \href {https://doi.org/10.1016/j.physletb.2019.134822} {\path{doi:10.1016/j.physletb.2019.134822}}.

\bibitem{Balachandran:1985fb}
A.~P. Balachandran, F.~Lizzi, V.~G.~J. Rodgers, A.~Stern, {Dibaryons as Chiral Solitons}, Nucl. Phys. B 256 (1985) 525--556.
\newblock \href {https://doi.org/10.1016/0550-3213(85)90407-9} {\path{doi:10.1016/0550-3213(85)90407-9}}.

\bibitem{Yost:1985mj}
S.~A. Yost, C.~R. Nappi, {The Mass of the $H$ Dibaryon in a Chiral Model}, Phys. Rev. D 32 (1985) 816.
\newblock \href {https://doi.org/10.1103/PhysRevD.32.816} {\path{doi:10.1103/PhysRevD.32.816}}.

\bibitem{Straub:1988mz}
U.~Straub, Z.-Y. Zhang, K.~Brauer, A.~Faessler, S.~B. Khadkikar, {Binding Energy of the Dihyperon in the Quark Cluster Model}, Phys. Lett. B 200 (1988) 241--245.
\newblock \href {https://doi.org/10.1016/0370-2693(88)90763-0} {\path{doi:10.1016/0370-2693(88)90763-0}}.

\bibitem{Kodama:1994np}
N.~Kodama, M.~Oka, T.~Hatsuda, {H dibaryon in the QCD sum rule}, Nucl. Phys. A 580 (1994) 445--454.
\newblock \href {http://arxiv.org/abs/hep-ph/9404221} {\path{arXiv:hep-ph/9404221}}, \href {https://doi.org/10.1016/0375-9474(94)90908-3} {\path{doi:10.1016/0375-9474(94)90908-3}}.

\bibitem{Nakamoto:1997gh}
C.~Nakamoto, Y.~Suzuki, Y.~Fujiwara, {Central force of baryon baryon interactions with S = -2 in the SU(6) quark model}, Prog. Theor. Phys. 97 (1997) 761--780.
\newblock \href {https://doi.org/10.1143/PTP.97.761} {\path{doi:10.1143/PTP.97.761}}.

\bibitem{Shen:2000qs}
P.-N. Shen, Z.-Y. Zhang, Y.-W. Yu, X.-Q. Yuan, S.~Yang, {Structure of H-dihyperon}, Chin. Phys. Lett. 17 (2000) 7--9.
\newblock \href {https://doi.org/10.1088/0256-307X/17/1/003} {\path{doi:10.1088/0256-307X/17/1/003}}.

\bibitem{Haidenbauer:2011ah}
J.~Haidenbauer, U.-G. Meissner, {To bind or not to bind: The H-dibaryon in light of chiral effective field theory}, Phys. Lett. B 706 (2011) 100--105.
\newblock \href {http://arxiv.org/abs/1109.3590} {\path{arXiv:1109.3590}}, \href {https://doi.org/10.1016/j.physletb.2011.10.070} {\path{doi:10.1016/j.physletb.2011.10.070}}.

\bibitem{Haidenbauer:2011za}
J.~Haidenbauer, U.~G. Meissner, {Exotic bound states of two baryons in light of chiral effective field theory}, Nucl. Phys. A 881 (2012) 44--61.
\newblock \href {http://arxiv.org/abs/1111.4069} {\path{arXiv:1111.4069}}, \href {https://doi.org/10.1016/j.nuclphysa.2012.01.021} {\path{doi:10.1016/j.nuclphysa.2012.01.021}}.

\bibitem{Marietti:2022jil}
D.~Marietti, A.~Pilloni, U.~Tamponi, {Production of loosely bound hadron molecules from bottomonium decays}, Phys. Rev. D 106~(9) (2022) 094040.
\newblock \href {http://arxiv.org/abs/2208.14185} {\path{arXiv:2208.14185}}, \href {https://doi.org/10.1103/PhysRevD.106.094040} {\path{doi:10.1103/PhysRevD.106.094040}}.

\bibitem{NPLQCD:2010ocs}
S.~R. Beane, et~al., {Evidence for a Bound H-dibaryon from Lattice QCD}, Phys. Rev. Lett. 106 (2011) 162001.
\newblock \href {http://arxiv.org/abs/1012.3812} {\path{arXiv:1012.3812}}, \href {https://doi.org/10.1103/PhysRevLett.106.162001} {\path{doi:10.1103/PhysRevLett.106.162001}}.

\bibitem{Inoue:2010hs}
T.~Inoue, N.~Ishii, S.~Aoki, T.~Doi, T.~Hatsuda, Y.~Ikeda, K.~Murano, H.~Nemura, K.~Sasaki, {Baryon-Baryon Interactions in the Flavor SU(3) Limit from Full QCD Simulations on the Lattice}, Prog. Theor. Phys. 124 (2010) 591--603.
\newblock \href {http://arxiv.org/abs/1007.3559} {\path{arXiv:1007.3559}}, \href {https://doi.org/10.1143/PTP.124.591} {\path{doi:10.1143/PTP.124.591}}.

\bibitem{Inoue:2010es}
T.~Inoue, N.~Ishii, S.~Aoki, T.~Doi, T.~Hatsuda, Y.~Ikeda, K.~Murano, H.~Nemura, K.~Sasaki, {Bound H-dibaryon in Flavor SU(3) Limit of Lattice QCD}, Phys. Rev. Lett. 106 (2011) 162002.
\newblock \href {http://arxiv.org/abs/1012.5928} {\path{arXiv:1012.5928}}, \href {https://doi.org/10.1103/PhysRevLett.106.162002} {\path{doi:10.1103/PhysRevLett.106.162002}}.

\bibitem{Hanlon:2018yfv}
A.~Hanlon, A.~Francis, J.~Green, P.~Junnarkar, H.~Wittig, {The $H$ dibaryon from lattice QCD with SU(3) flavor symmetry}, PoS LATTICE2018 (2018) 081.
\newblock \href {http://arxiv.org/abs/1810.13282} {\path{arXiv:1810.13282}}, \href {https://doi.org/10.22323/1.334.0081} {\path{doi:10.22323/1.334.0081}}.

\bibitem{Green:2021qol}
J.~R. Green, A.~D. Hanlon, P.~M. Junnarkar, H.~Wittig, {Weakly bound $H$ dibaryon from SU(3)-flavor-symmetric QCD}, Phys. Rev. Lett. 127~(24) (2021) 242003.
\newblock \href {http://arxiv.org/abs/2103.01054} {\path{arXiv:2103.01054}}, \href {https://doi.org/10.1103/PhysRevLett.127.242003} {\path{doi:10.1103/PhysRevLett.127.242003}}.

\bibitem{HALQCD:2019wsz}
K.~Sasaki, et~al., {$\Lambda\Lambda$ and N$\Xi$ interactions from lattice QCD near the physical point}, Nucl. Phys. A 998 (2020) 121737.
\newblock \href {http://arxiv.org/abs/1912.08630} {\path{arXiv:1912.08630}}, \href {https://doi.org/10.1016/j.nuclphysa.2020.121737} {\path{doi:10.1016/j.nuclphysa.2020.121737}}.

\bibitem{Briceno:2015tza}
R.~A. Brice{\~n}o, M.~T. Hansen, {Relativistic, model-independent, multichannel $2\to 2$ transition amplitudes in a finite volume}, Phys. Rev. D 94~(1) (2016) 013008.
\newblock \href {http://arxiv.org/abs/1509.08507} {\path{arXiv:1509.08507}}, \href {https://doi.org/10.1103/PhysRevD.94.013008} {\path{doi:10.1103/PhysRevD.94.013008}}.

\bibitem{Aitchison:1972ay}
I.~J.~R. Aitchison, {K-MATRIX FORMALISM FOR OVERLAPPING RESONANCES}, Nucl. Phys. A 189 (1972) 417--423.
\newblock \href {https://doi.org/10.1016/0375-9474(72)90305-3} {\path{doi:10.1016/0375-9474(72)90305-3}}.

\bibitem{SupplementalMaterial}
Supplemental material, supplementary material submitted with this article.

\bibitem{Castillejo:1955ed}
L.~Castillejo, R.~H. Dalitz, F.~J. Dyson, {Low's scattering equation for the charged and neutral scalar theories}, Phys. Rev. 101 (1956) 453--458.
\newblock \href {https://doi.org/10.1103/PhysRev.101.453} {\path{doi:10.1103/PhysRev.101.453}}.

\bibitem{Bruno:2014jqa}
M.~Bruno, et~al., {Simulation of QCD with N$_{f} =$ 2 $+$ 1 flavors of non-perturbatively improved Wilson fermions}, JHEP 02 (2015) 043.
\newblock \href {http://arxiv.org/abs/1411.3982} {\path{arXiv:1411.3982}}, \href {https://doi.org/10.1007/JHEP02(2015)043} {\path{doi:10.1007/JHEP02(2015)043}}.

\bibitem{Bruno:2016plf}
M.~Bruno, T.~Korzec, S.~Schaefer, {Setting the scale for the CLS $2 + 1$ flavor ensembles}, Phys. Rev. D 95~(7) (2017) 074504.
\newblock \href {http://arxiv.org/abs/1608.08900} {\path{arXiv:1608.08900}}, \href {https://doi.org/10.1103/PhysRevD.95.074504} {\path{doi:10.1103/PhysRevD.95.074504}}.

\bibitem{HadronSpectrum:2009krc}
M.~Peardon, J.~Bulava, J.~Foley, C.~Morningstar, J.~Dudek, R.~G. Edwards, B.~Joo, H.-W. Lin, D.~G. Richards, K.~J. Juge, {A Novel quark-field creation operator construction for hadronic physics in lattice QCD}, Phys. Rev. D 80 (2009) 054506.
\newblock \href {http://arxiv.org/abs/0905.2160} {\path{arXiv:0905.2160}}, \href {https://doi.org/10.1103/PhysRevD.80.054506} {\path{doi:10.1103/PhysRevD.80.054506}}.

\bibitem{Luscher:1990ck}
M.~Luscher, U.~Wolff, {How to Calculate the Elastic Scattering Matrix in Two-dimensional Quantum Field Theories by Numerical Simulation}, Nucl. Phys. B 339 (1990) 222--252.
\newblock \href {https://doi.org/10.1016/0550-3213(90)90540-T} {\path{doi:10.1016/0550-3213(90)90540-T}}.

\bibitem{Blossier:2009kd}
B.~Blossier, M.~Della~Morte, G.~von Hippel, T.~Mendes, R.~Sommer, {On the generalized eigenvalue method for energies and matrix elements in lattice field theory}, JHEP 04 (2009) 094.
\newblock \href {http://arxiv.org/abs/0902.1265} {\path{arXiv:0902.1265}}, \href {https://doi.org/10.1088/1126-6708/2009/04/094} {\path{doi:10.1088/1126-6708/2009/04/094}}.

\bibitem{Briceno:2018aml}
R.~A. Brice{\~n}o, M.~T. Hansen, S.~R. Sharpe, {Three-particle systems with resonant subprocesses in a finite volume}, Phys. Rev. D 99~(1) (2019) 014516.
\newblock \href {http://arxiv.org/abs/1810.01429} {\path{arXiv:1810.01429}}, \href {https://doi.org/10.1103/PhysRevD.99.014516} {\path{doi:10.1103/PhysRevD.99.014516}}.

\bibitem{Hansen:2024cai}
M.~T. Hansen, T.~Peterken, {Discretization effects in finite-volume $2\to2$ scattering} (8 2024).
\newblock \href {http://arxiv.org/abs/2408.07062} {\path{arXiv:2408.07062}}.

\bibitem{James:1975dr}
F.~James, M.~Roos, {Minuit: A System for Function Minimization and Analysis of the Parameter Errors and Correlations}, Comput. Phys. Commun. 10 (1975) 343--367.
\newblock \href {https://doi.org/10.1016/0010-4655(75)90039-9} {\path{doi:10.1016/0010-4655(75)90039-9}}.

\bibitem{iminuit}
H.~Dembinski, P.~O. et~al., \href{https://doi.org/10.5281/zenodo.3949207}{scikit-hep/iminuit} (Dec 2020).
\newblock \href {https://doi.org/10.5281/zenodo.3949207} {\path{doi:10.5281/zenodo.3949207}}.
\newline\urlprefix\url{https://doi.org/10.5281/zenodo.3949207}

\bibitem{Oller:1998zr}
J.~A. Oller, E.~Oset, {N/D description of two meson amplitudes and chiral symmetry}, Phys. Rev. D 60 (1999) 074023.
\newblock \href {http://arxiv.org/abs/hep-ph/9809337} {\path{arXiv:hep-ph/9809337}}, \href {https://doi.org/10.1103/PhysRevD.60.074023} {\path{doi:10.1103/PhysRevD.60.074023}}.

\bibitem{Wilson:2014cna}
D.~J. Wilson, J.~J. Dudek, R.~G. Edwards, C.~E. Thomas, {Resonances in coupled $\pi K, \eta K$ scattering from lattice QCD}, Phys. Rev. D 91~(5) (2015) 054008.
\newblock \href {http://arxiv.org/abs/1411.2004} {\path{arXiv:1411.2004}}, \href {https://doi.org/10.1103/PhysRevD.91.054008} {\path{doi:10.1103/PhysRevD.91.054008}}.

\bibitem{Du:2024gzw}
M.-L. Du, F.-K. Guo, B.~Wu, {Effective-Range Expansion with a Long-Range Force}, Phys. Rev. Lett. 135~(1) (2025) 011903.
\newblock \href {http://arxiv.org/abs/2408.09375} {\path{arXiv:2408.09375}}, \href {https://doi.org/10.1103/cwdt-dj6z} {\path{doi:10.1103/cwdt-dj6z}}.

\bibitem{Wang:2026ups}
W.-J. Wang, B.~Wu, M.-L. Du, F.-K. Guo, {Effective Range Expansion with the Left-Hand Cut: Higher Order Improvements} (1 2026).
\newblock \href {http://arxiv.org/abs/2601.04989} {\path{arXiv:2601.04989}}.

\bibitem{Liu:2026xrk}
B.-Y. Liu, B.~Wu, J.-W. Fu, M.-L. Du, F.-K. Guo, U.-G. Mei{\ss}ner, {Extraction of the pion-nucleon coupling constant using the effective-range expansion with the left-hand cut} (2 2026).
\newblock \href {http://arxiv.org/abs/2602.03115} {\path{arXiv:2602.03115}}.

\bibitem{Oller:2018zts}
J.~A. Oller, D.~R. Entem, {The exact discontinuity of a partial wave along the left-hand cut and the exact $N/D$ method in non-relativistic scattering}, Annals Phys. 411 (2019) 167965.
\newblock \href {http://arxiv.org/abs/1810.12242} {\path{arXiv:1810.12242}}, \href {https://doi.org/10.1016/j.aop.2019.167965} {\path{doi:10.1016/j.aop.2019.167965}}.

\bibitem{Baru:2021ldu}
V.~Baru, X.-K. Dong, M.-L. Du, A.~Filin, F.-K. Guo, C.~Hanhart, A.~Nefediev, J.~Nieves, Q.~Wang, {Effective range expansion for narrow near-threshold resonances}, Phys. Lett. B 833 (2022) 137290.
\newblock \href {http://arxiv.org/abs/2110.07484} {\path{arXiv:2110.07484}}, \href {https://doi.org/10.1016/j.physletb.2022.137290} {\path{doi:10.1016/j.physletb.2022.137290}}.

\bibitem{Garcia-Martin:2011nna}
R.~Garcia-Martin, R.~Kaminski, J.~R. Pelaez, J.~Ruiz~de Elvira, {Precise determination of the f0(600) and f0(980) pole parameters from a dispersive data analysis}, Phys. Rev. Lett. 107 (2011) 072001.
\newblock \href {http://arxiv.org/abs/1107.1635} {\path{arXiv:1107.1635}}, \href {https://doi.org/10.1103/PhysRevLett.107.072001} {\path{doi:10.1103/PhysRevLett.107.072001}}.

\bibitem{Pelaez:2020gnd}
J.~R. Pel{\'a}ez, A.~Rodas, {Dispersive {\ensuremath{\pi}}K{\textrightarrow}{\ensuremath{\pi}}K and {\ensuremath{\pi}}{\ensuremath{\pi}}{\textrightarrow}$K\bar{K}$ amplitudes from scattering data, threshold parameters, and the lightest strange resonance {\ensuremath{\kappa}} or K0{\ensuremath{*}}(700)}, Phys. Rept. 969 (2022) 1--126.
\newblock \href {http://arxiv.org/abs/2010.11222} {\path{arXiv:2010.11222}}, \href {https://doi.org/10.1016/j.physrep.2022.03.004} {\path{doi:10.1016/j.physrep.2022.03.004}}.

\bibitem{Weinberg:1962hj}
S.~Weinberg, {Elementary particle theory of composite particles}, Phys. Rev. 130 (1963) 776--783.
\newblock \href {https://doi.org/10.1103/PhysRev.130.776} {\path{doi:10.1103/PhysRev.130.776}}.

\bibitem{Weinberg:1965zz}
S.~Weinberg, {Evidence That the Deuteron Is Not an Elementary Particle}, Phys. Rev. 137 (1965) B672--B678.
\newblock \href {https://doi.org/10.1103/PhysRev.137.B672} {\path{doi:10.1103/PhysRev.137.B672}}.

\bibitem{Baru:2003qq}
V.~Baru, J.~Haidenbauer, C.~Hanhart, Y.~Kalashnikova, A.~E. Kudryavtsev, {Evidence that the a(0)(980) and f(0)(980) are not elementary particles}, Phys. Lett. B 586 (2004) 53--61.
\newblock \href {http://arxiv.org/abs/hep-ph/0308129} {\path{arXiv:hep-ph/0308129}}, \href {https://doi.org/10.1016/j.physletb.2004.01.088} {\path{doi:10.1016/j.physletb.2004.01.088}}.

\bibitem{Matuschek:2020gqe}
I.~Matuschek, V.~Baru, F.-K. Guo, C.~Hanhart, {On the nature of near-threshold bound and virtual states}, Eur. Phys. J. A 57~(3) (2021) 101.
\newblock \href {http://arxiv.org/abs/2007.05329} {\path{arXiv:2007.05329}}, \href {https://doi.org/10.1140/epja/s10050-021-00413-y} {\path{doi:10.1140/epja/s10050-021-00413-y}}.

\bibitem{Li:2021cue}
Y.~Li, F.-K. Guo, J.-Y. Pang, J.-J. Wu, {Generalization of Weinberg{\textquoteright}s compositeness relations}, Phys. Rev. D 105~(7) (2022) L071502.
\newblock \href {http://arxiv.org/abs/2110.02766} {\path{arXiv:2110.02766}}, \href {https://doi.org/10.1103/PhysRevD.105.L071502} {\path{doi:10.1103/PhysRevD.105.L071502}}.

\bibitem{Esposito:2021vhu}
A.~Esposito, L.~Maiani, A.~Pilloni, A.~D. Polosa, V.~Riquer, {From the line shape of the X(3872) to its structure}, Phys. Rev. D 105~(3) (2022) L031503.
\newblock \href {http://arxiv.org/abs/2108.11413} {\path{arXiv:2108.11413}}, \href {https://doi.org/10.1103/PhysRevD.105.L031503} {\path{doi:10.1103/PhysRevD.105.L031503}}.

\bibitem{Albaladejo:2022sux}
M.~Albaladejo, J.~Nieves, {Compositeness of S-wave weakly-bound states from next-to-leading order Weinberg{\textquoteright}s relations}, Eur. Phys. J. C 82~(8) (2022) 724.
\newblock \href {http://arxiv.org/abs/2203.04864} {\path{arXiv:2203.04864}}, \href {https://doi.org/10.1140/epjc/s10052-022-10695-1} {\path{doi:10.1140/epjc/s10052-022-10695-1}}.

\end{thebibliography}
\end{document}
\fi

\end{document}